\documentclass[preprint,aps,11pt,showpacs,nofootinbib,tightenlines]{revtex4}
\usepackage{amsmath}
\usepackage{amssymb}
\usepackage{epsfig}
\usepackage{graphicx}
\begin{document}

\newcommand{\beq}{\begin{eqnarray}}
\newcommand{\eeq}{\end{eqnarray}}
\newcommand{\non}{\nonumber\\ }

\newcommand{\acp}{{\cal A}_{CP}}
\newcommand{\etap}{\eta^{(\prime)} }
\newcommand{\etapr}{\eta^\prime }
\newcommand{\jpsi}{ J/\Psi }
\newcommand{\kst} {K^{*0}}
\newcommand{\kstb}{\overline{K}^{*0}}

\newcommand{\psl}{ P \hspace{-2.8truemm}/ }
\newcommand{\nsl}{ n \hspace{-2.2truemm}/ }
\newcommand{\vsl}{ v \hspace{-2.2truemm}/ }
\newcommand{\epsl}{\epsilon \hspace{-1.8truemm}/\,  }

\def \epjc{ Eur. Phys. J. C }
\def \jpg{  J. Phys. G }
\def \npb{  Nucl. Phys. B }
\def \plb{  Phys. Lett. B }
\def \pr{  Phys. Rep. }
\def \prd{  Phys. Rev. D }
\def \prl{  Phys. Rev. Lett.  }
\def \zpc{  Z. Phys. C  }
\def \jhep{ J. High Energy Phys.  }
\def \ijmpa { Int. J. Mod. Phys. A }
\def \cpc{ Chin. Phys. C }
\def \ctp{ Commun. Theor. Phys. }
\def \rmp{ Rev. Mod. Phys. }

\title{ Light scalar mesons and charmless hadronic $B_c \to SP, SV$ decays in the perturbative QCD approach }
\author{Xin~Liu\footnote{liuxin.physics@gmail.com} and
 Zhen-Jun~Xiao\footnote{xiaozhenjun@njnu.edu.cn}}
\affiliation{ Department of Physics and Institute of Theoretical
Physics, Nanjing Normal University,\\
Nanjing, Jiangsu 210046, People's Republic of China }
\date{\today}
\begin{abstract}
The scalar productions
in heavy meson decays can provide
a good platform to study not only heavy flavor physics but also
their own physical properties
in a dramatically different way.
In this work, based on the assumption of two-quark structure of the scalars,
the charmless hadronic $B_c \to SP,
SV$ decays(here, $S$, $P$, and $V$ denote the light scalar, pseudoscalar, and vector mesons, respectively)
are investigated by employing
the perturbative QCD(pQCD) factorization approach. In the standard model
all these considered $B_c$ meson decays can
only occur through the annihilation diagrams.
From our numerical evaluations and phenomenological analysis, we
find that
(a) the pQCD predictions for the {\it CP}-averaged branching
ratios(BRs) of the considered $B_c$ decays vary in the range of
$10^{-5}$ to $10^{-8}$, which will be tested in the ongoing LHCb and forthcoming Super-B experiments,
 while the {\it CP}-violating asymmetries for these modes
are absent naturally in the standard model because
only one type tree operator is involved;
(b) for $B_c \to SP, SV$ decays, the BRs
of $\Delta S= 0$ processes are basically much larger than those of
$\Delta S =1$ ones as generally expected because the different
Cabibbo-Kobayashi-Maskawa(CKM) factors are involved;
(c) analogous
to $B \to K^* \eta^{(\prime)}$ decays,
 $Br(B_c \to \kappa^+ \eta) \sim 5 \times Br(B_c \to \kappa^+ \eta^\prime)$ in the pQCD approach, which can be
understood by the constructive and destructive interference
between the $\eta_q$ and $\eta_s$ contributions to the $B_c \to \kappa^+
\eta$ and $B_c \to \kappa^+ \eta^\prime$ decays, however, $Br(B_c \to K_0^*(1430) \eta)$ is approximately equal to
$Br(B_c \to K_0^*(1430) \eta')$ in both scenarios because the factorizable contributions from $\eta_s$ term
play the dominant role in the considered two channels;
(d) if $a_0(980)$ and $\kappa$ are the $q\bar q$ bound states, the pQCD predicted
BRs for $B_c \to a_0(980) (\pi, \rho)$ and $B_c \to \kappa K^{(*)}$ decays will be in the range of
$10^{-6} \sim 10^{-5}$, which are within the reach of the LHCb experiments and could be measured in the near future;
and
(e) for the $a_0(1450)$ and $K_0^*(1430)$ channels, the BRs for $B_c \to a_0(1450) (\pi, \rho)$
and $B_c \to K_0^*(1430) K^{(*)}$ modes in the pQCD approach are found to be
$(5 \sim 47) \times 10^{-6}$
and $(0.7 \sim 36) \times 10^{-6}$, respectively. A measurement of them at the predicted level will
favor the $q\bar q$ structure and help understand the physical properties of the scalars and the involved QCD
dynamics in the modes, especially the reliability of the pQCD approach to these
$B_c$ meson decays.

\end{abstract}

\pacs{13.25.Hw, 12.38.Bx, 14.40.Nd}

\maketitle

\section{Introduction}

The scalar mesons are especially important to understand because
they have the same quantum numbers as the vacuum ($J^{PC} =
0^{++}$). Great efforts have been made by the physicists on both
experimental and theoretical aspects to understand the inner
structure of the scalars
but it
is well-known that
the underlying structure of them
is not yet well established
(for a review,
see e.g.~\cite{Amsler08:pdg,Godfrey99:scalar,Close02:scalar}). Up
to now, many different possible solutions to
the scalars have been proposed such as
$\bar qq$, $\bar q\bar qqq$, meson-meson bound states or even
supplemented with a scalar glueball.
More likely, they are not made of one simple component but are the superpositions of
these contents. The different scenarios tend to give very
different predictions on the production and decay of the scalar
mesons which are helpful to determine the dominant component.

The first charmless $B$ decay into a scalar meson, i.e., $B \to f_0(980) K$, was measured
by Belle~\cite{belle02:f98} in 2002 (updated in~\cite{belle05:f98})
and subsequently confirmed by BaBar~\cite{babar04:f98} in 2004. After that these two $B$ factories
operated at KEK and SLAC respectively have found many decay channels with the scalars as one of the
productions in $B$ meson decays~\cite{Amsler08:pdg,Barberio08:hfag}. These measurements should provide information on the nature of the
scalar mesons. It is of enough reasons to believe that as
a different unique insight to the internal structure of the scalars, the heavy
$B$ meson decaying into scalar mesons can provide a good place to explore
their physical properties.

Recently, the production of scalar mesons with $q\bar q$ structure in the two-body
charmless $B$ decays have been intensively studied in
Refs.~\cite{Chen:b2scalar,Cheng:b2scalar,Cheng06:B2SP,Wang:b2scalar,Xiao:b2scalar}
theoretically, in which many predictions are within the reach of the current $B$ factory experiments
and to be examined in the near future.
It is hoped that through the study of $B \to SP,
SV$(Here, $S$, $P$, and $V$ are the light scalar, pseudoscalar, and vector
mesons, respectively.) decays, old puzzles related to the internal
structure and related parameters, e.g., the masses and widths, of
light scalar mesons can receive new understanding.

Experimentally, the Large Hadron Collider (LHC) experiment at CERN is running now,
where the
$B_c$ meson could be produced abundantly. Motivated
by the
forthcoming large number of $B_c$ production and decay events
in the ongoing LHCb experiments, the scalar meson spectrum
would become one of the most interesting topics for both experimental
and theoretical studies in the near future. At that time, more and
more channels with scalar mesons will be opened and got stringent
tests from
the experiments, which will help us to further explore
the nature of the scalars. On the other hand,
for $B_c$ meson, one can
study the two heavy flavors b and c in a meson simultaneously. The
$B_c$ meson decays may also provide windows for studying the perturbative
and nonperturbative QCD, final state interactions, testing the
predictions of the standard model(SM), and can shed light on new
physics
scenarios beyond the SM~\cite{nb04:bcre}.

Inspired by the above observations, in this work, we therefore will focus on the two-body
charmless hadronic decays $B_c \to SP, SV$, which can only occur through the weak
annihilation diagrams. The size of annihilation contributions is
an important issue in $B$ physics for many years.
The importance of
annihilation contributions has already been tested in the previous
predictions of branching ratios of pure annihilation $B \to D_s K$
decays~\cite{Lu03:dsk}, direct CP asymmetries of $B^0 \to
\pi^+\pi^-$, $K^+\pi^-$
decays~\cite{Keum01:kpi,Lu01:pipi,Hong06:direct} and in the
explanation of $B\to \phi K^*$ polarization
problem~\cite{Li05:kphi,Gritsan07:kphi} though there still exist much different viewpoints\footnote{Recently, the authors announced
in Ref.~\cite{Stewart08:anni-scet} that the annihilation contributions in charmless hadronic $B$ decays
are real and small in the soft-collinear effective theory~\cite{scet} at leading power.
While the authors in another work~\cite{Chay08:complexanni} discussed that they may be the almost imaginary contributions, which
can generate a sizable strong phase.
 This discrepancy between these two approaches/methods needs to be clarified definitely by the experiments in the future.}. The two-body $B$ decays
into the final states with one scalar meson may suffer from large
weak annihilation contributions, which have been analyzed in
Refs.~\cite{Cheng06:B2SP,Wang:b2scalar,Xiao:b2scalar} preliminary. Thus it is very interesting
to explore the size of annihilation contributions in
these considered $B_c \to SP, SV$ channels, which will also be
helpful to investigate the annihilated decay mechanism and the physical properties of the scalars.

In this paper, we will study
the {\it CP}-averaged branching
ratios(BRs) of charmless hadronic $B_c \to SP, SV$ decays by employing the low energy
effective Hamiltonian~\cite{Buras96:weak} and the perturbative
QCD(pQCD) factorization
approach~\cite{Keum01:kpi,Lu01:pipi,Li03:ppnp}. By keeping the
transverse momentum $k_T$ of the quarks, the pQCD approach is free
of endpoint singularity and the Sudakov formalism makes it more
self-consistent.
Rather different from the QCD factorization approach~\cite{qcdf} and
 soft-collinear effective theory,
 the pQCD approach can be used to calculate
the annihilation diagrams straightforwardly~\cite{Li09:review}, as have been done for example in
Refs.~\cite{Keum01:kpi,Lu01:pipi,Lu03:dsk,Li05:kphi,xiao06,Hong06:direct,
Wang:b2scalar,Xiao:b2scalar,Ali07:bsnd,xiao08:keta,Xiao1:bcdecay}.

The paper is organized as follows. In Sec.~\ref{sec:1}, we present
a brief review of light scalar mesons and the formalism of pQCD approach.
The wave functions and distribution
amplitudes for heavy $B_c$ and light scalar, pseudoscalar, and vector
mesons are also given here.
Then we perform the perturbative calculations for the considered
$B_c \to SP, SV$ decay channels with pQCD approach in Sec.~\ref{sec:2}.
The analytic formulas of the decay amplitudes for all the considered
 modes are also collected in this section.
The numerical results and phenomenological analysis are given in
Sec.~\ref{sec:3}. Finally, Sec.~\ref{sec:sum} contains the main
conclusions and a short summary.

\section{Light scalar mesons, formalism, and wave functions}\label{sec:1}

\subsection{Light scalar mesons}

Up to now, the people have discovered many scalar states
experimentally but know little about their underlying
structures, which are 
not well established theoretically yet
(for a review, see Refs.~\cite{Amsler08:pdg,Godfrey99:scalar,Close02:scalar}). According to the
meson particle collected by the Particle Data
Group~\cite{Amsler08:pdg}, the light scalar mesons below or near 1
GeV, including $a_0(980)$, $K^*_0(800)({\rm or}~ \kappa)$,
$f_0(600)({\rm or}~ \sigma)$, and $f_0(980)$, are usually viewed
to form an SU(3) flavor nonet; while scalar mesons around 1.5 GeV,
including $a_0(1450)$, $K^*_0(1430)$, $f_0(1370)$, and
$f_0(1500)/f_0(1710)$, form another nonet\footnote{For the sake of
simplicity, we will use $a_0$ and $f_0$ to denote $a_0(980)$ and $f_0(980)$, respectively, unless otherwise stated.
We will also adopt the forms $a$, $K_0^*$, $f$, and
$f'$ to denote the scalar mesons $a_0(980)$ and $a_0(1450)$,
$K_0^*(800)$ and $K_0^*(1430)$, $f_0(600)$ and $f_0(1370)$, and
$f_0(980)$ and $f_0(1500)/f_0(1710)$ correspondingly in the
following sections, unless otherwise stated. }.

Recently, Cheng, Chua, and Yang~\cite{Cheng06:B2SP} proposed two possible
scenarios to describe these light scalar mesons in the QCD sum rule
method:
\begin{enumerate}

\item
In scenario 1(S1), the scalar mesons in the former
nonet are
treated as the lowest lying states, and in the latter
one as the
corresponding first excited states, respectively. Based on the
naive two-quark model, the flavor structure
of the light
scalar mesons in S1 read
 \beq
 \sigma&=& \frac{1}{\sqrt{2}}(u\bar u+d\bar d)\;,\;\;\;\;\; f_0=s\bar s\;,\non
 a_0^+&=& u\bar d\;,\;\;  a_0^0=\frac{1}{\sqrt{2}}(u\bar u-d\bar d)\;,\;\;  a_0^-=d\bar
 u\;, \non
 \kappa^+&=&u\bar s\;,\;\;\;\; \kappa^0=d\bar s\;,\;\;\;\; \bar\kappa^0=s\bar
 d\;,\;\;\;\; \kappa^-=s\bar u\;.
 \eeq
Here, it is assumed that the lightest $\sigma$ and heaviest $f_0$ in the lighter scalar nonet has the ideal
mixing. But various experimental data
indicate
that $f_0$ should not have the pure $s\bar s$ component and the isoscalars $\sigma$ and $f_0$ must
have a mixing of $f_0^q$ and $f_0^s$~\cite{Cheng06:B2SP}, which is analogous to $\eta-\eta'$ mixing system,
\beq
\left ( \begin{array}{l} \sigma\\ f_0 \\ \end{array} \right ) =
\left ( \begin{array}{rr}
\cos\theta_0 & -\sin\theta_0\\ \sin\theta_0& \cos\theta_0\\ \end{array} \right )
\left ( \begin{array}{l} f_0^q\\ f_0^s\\ \end{array} \right )\;,
\label{eq:f0mixing}
\eeq
with $f_0^q= (\bar u u +\bar d d)/\sqrt{2}$ and $f_0^s = \bar s s$,
 where $\theta_0$ is the mixing angle between $\sigma$ and $f_0$.
Many works
have been made to explore the mixing angle $\theta_0$~\cite{Mixangle:theta0}:
$\theta_0$ lies in the ranges of $25^\circ<\theta_0<40^\circ$
and $140^\circ<\theta_0< 165^\circ$. But the fact that $\theta_0$
tends to be not a unique value, which indicates that
$\sigma$ and $f_0$ may not be purely $q\bar q$ states.

While for the mixing of the isosinglet scalar mesons $f_0(1370)$, $f_0(1500)$, and $f_0(1710)$,
which have been
discussed in detail in the literatures (See Ref.~\cite{Cheng06:glueball} and references therein).
In this work, we will adopt the
mixing mechanism as given in Ref.~\cite{Cheng06:glueball},
 \beq
\left(\begin{array}{l}  f_0(1370) \\ f_0(1500) \\ f_0(1710) \\ \end{array}\right)=
\left(\begin{array}{rrr}  0.78 & 0.51 & -0.36 \\
                         -0.54 & 0.84 & 0.03 \\
                          0.32 & 0.18 & 0.93 \\
                                                               \end{array}\right)
\left(\begin{array}{l}  f_0^q \\ f_0^s \\ f_0^G  \\ \end{array}\right).
\label{eq:f0mixing1}
 \eeq
As discussed in~\cite{Cheng06:glueball}, it is evident that $f_0(1370)$ and $f_0(1500)$
mainly consists of $f_0^q$ and $f_0^s$, just with small or tiny glueball components, however,
$f_0(1710)$ is composed primarily of the scalar glueball, i.e., $f_0^G$.
We will therefore only take the scalar mesons $f_0(1370)$ and $f_0(1500)$ into account
in the present work, and leave the contribution
from scalar glueball content for future study.

\item In scenario 2(S2) that the scalar mesons in the latter nonet
are the lowest lying resonances and the corresponding first
excited states lie between $(2.0\sim 2.3)$~GeV. S2 corresponds to
the case that light scalar mesons below or near 1
GeV are four-quark bound states,
while all scalar mesons are made of two quarks in S1.
 In order to give quantitative predictions, since we do not know how to deal with the four-quark states in the
factorization approach presently, we here just consider the evaluations on
the scalar mesons with $q\bar q$ structure
in S2.

\end{enumerate}

In short, we will investigate these light scalar mesons in the
pure annihilation $B_c \to SP, SV$ decays with
the assumption of two-quark structure proposed in the above two possible scenarios.

\subsection{Formalism of pQCD approach}

Since the b quark is rather heavy, we work in the frame with the
$B_c$ meson at rest, i.e., with the $B_c$ meson momentum
$P_1=\frac{m_{B_c}}{\sqrt{2}}(1,1,{\bf 0}_T)$ in the light-cone
coordinates. For the charmless hadronic $B_c \to M_2 M_3$
\footnote{For the sake of simplicity, we will use $M_2$ and $M_3$ to denote the two
final state light mesons respectively, unless otherwise stated. }
decays, assuming that the $M_2$ ($M_3$) meson moves in the plus
(minus) $z$ direction carrying the momentum $P_2$ ($P_3$) and the longitudinal
polarization vector $\epsilon_2^L$ ($\epsilon_3^L$)(if $M_{2(3)}$ is
the vector meson). Then the two final state meson momenta can be
written as
\beq
     P_2 =\frac{m_{B_c}}{\sqrt{2}} (1-r_3^2,r_2^2,{\bf 0}_T), \quad
     P_3 =\frac{m_{B_c}}{\sqrt{2}} (r_3^2,1-r_2^2,{\bf 0}_T),
\eeq
respectively, where $r_2=m_{M_2}/m_{B_c}$ and
$r_3=m_{M_3}/m_{B_c}$. When $M_2~ {\rm or}~ M_3$ is a vector meson, the
longitudinal polarization vector, $\epsilon_2^L$ or
$\epsilon_3^L$, can be given by
\beq
\epsilon_2^L =\frac{m_{B_c}}{\sqrt{2}m_{M_2}} (1-r_3^2, -r_2^2,{\bf 0}_T), \;\;\;\;\;  {\rm or }\;\;\;\;\
\epsilon_3^L = \frac{m_{B_c}}{\sqrt{2}m_{M_3}} (-r_3^2, 1-r_2^2,{\bf0}_T).
\eeq
Putting the (light-)
 quark momenta in $B_c$, $M_2$ and $M_3$ mesons as $k_1$,
$k_2$, and $k_3$, respectively, we can choose
\beq
k_1 = (x_1P_1^+,0,{\bf k}_{1T}), \quad k_2 = (x_2 P_2^+,0,{\bf k}_{2T}), \quad
k_3 = (0, x_3 P_3^-,{\bf k}_{3T}).
\eeq
Then, for $B_c \to M_2 M_3$
decays, the
integration over $k_1^-$, $k_2^-$, and $k_3^+$
will conceptually lead to the decay amplitudes in the pQCD approach,
\beq
{\cal A}(B_c \to M_2 M_3) &\sim &\int\!\! d x_1 d x_2 d x_3 b_1
d b_1 b_2 d b_2 b_3 d b_3 \non && \cdot \mathrm{Tr} \left [ C(t)
\Phi_{B_c}(x_1,b_1) \Phi_{M_2}(x_2,b_2) \Phi_{M_3}(x_3, b_3) H(x_i,
b_i, t) S_t(x_i)\, e^{-S(t)} \right ]\;,
\label{eq:a2}
\eeq
where $b_i$ is the conjugate space coordinate of $k_{iT}$, and $t$ is the
largest energy scale in function $H(x_i,b_i,t)$. The large
logarithms $\ln (m_W/t)$ are included in the Wilson coefficients
$C(t)$. The large double logarithms ($\ln^2 x_i$) are summed by the
threshold resummation~\cite{Li02:resum}, and they lead to
$S_t(x_i)$ which smears the end-point singularities on $x_i$. The
last term, $e^{-S(t)}$, is the Sudakov form factor which suppresses
the soft dynamics effectively~\cite{Li98:soft}. Thus it makes the
perturbative calculation of the hard part $H$ applicable at
intermediate scale, i.e., $m_{B_c}$ scale. We will calculate
analytically the function $H(x_i,b_i,t)$ for the considered decays
at leading order(LO) in $\alpha_s$ expansion and give the convoluted
amplitudes in next section.

For these considered decays, the related weak effective
Hamiltonian $H_{{\rm eff}}$~\cite{Buras96:weak} can be written as
\beq
\label{eq:heff} H_{{\rm eff}} = \frac{G_{F}}
{\sqrt{2}} \, \left[ V_{cb}^* V_{uD} \left (C_1(\mu) O_1(\mu) +
C_2(\mu) O_2(\mu) \right) \right] \;\label{heff} ,
\eeq
with the local four-quark tree operators $O_{1,2}$
\beq
O_1 &=& \bar u_\beta \gamma^\mu (1-\gamma_5) D_\alpha \bar c_\beta \gamma^\mu
(1- \gamma_5) b_\alpha \; , \non
O_2 &=& \bar u_\beta \gamma^\mu (1- \gamma_5) D_\beta
\bar c_\alpha \gamma^\mu (1- \gamma_5) b_\alpha \; ,
\eeq
where $V_{cb}, V_{uD}$ are the Cabibbo-Kobayashi-Maskawa (CKM) matrix
elements, "$D$" denotes the light down quark $d$ or $s$ and
$C_i(\mu)$ are Wilson coefficients at the renormalization scale
$\mu$. For the Wilson coefficients $C_{1,2}(\mu)$, we will also
use the leading order expressions, although the
next-to-leading order calculations already exist in the
literature~\cite{Buras96:weak}. This is the consistent way to
cancel the explicit $\mu$ dependence in the theoretical formulae.
For the renormalization group evolution of the Wilson coefficients
from higher scale to lower scale, we use the formulas as given in
Ref.~\cite{Lu01:pipi} directly.

\subsection{Wave functions and distribution amplitudes}\label{ssec:wf}

In order to calculate the decay amplitude, we should choose the
proper wave function of the heavy $B_c$ meson. In
principle there are two Lorentz structures in the $B_{q}(q=u,d,s)$ or
$B_c$ meson wave function. One should consider both of them in
calculations. However, since the contribution induced by one
Lorentz structure is numerically small~\cite{luy03:form,Ali07:bsnd}
and can be neglected approximately, we only consider the
contribution from the first Lorentz structure.
\beq \Phi_{B_c} (x)
&=& \frac{i}{\sqrt{2 N_c}}\left[ (\psl  + M_{B_c}) \gamma_5
\phi_{B_c}(x) \right]_{\alpha\beta}\;.
\eeq
Since $B_c$ meson
consists of two heavy quarks and $m_{B_c} \simeq m_b+m_c$, the
distribution amplitude $\phi_{B_c}$ would be close to
$\delta(x-m_c/m_{B_c})$ in the non-relativistic limit. We
therefore adopt the non-relativistic approximation form of
$\phi_{B_c}$ as~\cite{CDL}, \beq \phi_{B_c}(x) &=&
\frac{f_{B_c}}{2 \sqrt{2 N_c}} \delta (x- m_c/m_{B_c})\;, \eeq
where $f_{B_c}$ and $N_c$ are the decay constant of $B_c$ meson
and the color number, respectively.

The wave function for the scalar meson$(S)$ can generally be
defined as,
\beq
\Phi_S(x) &=& \frac{i}{\sqrt{2 N_c}}
\left\{\psl \phi_S(x)+ m_S \phi_S^S(x) + m_S (\nsl \vsl -1)
\phi_S^T(x)\right\}_{\alpha\beta}\;,
\eeq
where $\phi_S$ and $\phi_S^{S,T}$, and $m_S$ are the leading twist and twist-3 distribution amplitudes, and mass
of the scalar meson, respectively, while $x$ denotes the momentum
fraction carried by quark in the meson, and $n=(1,0,{\bf 0}_T)$
and $v=(0,1,{\bf 0}_T)$ are dimensionless light-like unit vectors.

In general, the leading twist light-cone distribution amplitude
$\phi_{S}(x,\mu)$ can be expanded as the Gegenbauer
polynomials~\cite{Cheng06:B2SP,Li09:B2Sfm}: \beq
\phi_{S}(x,\mu)&=&\frac{3}{\sqrt{2N_c}}x(1-x)\biggl\{f_{S}(\mu)+\bar
f_{S}(\mu)\sum_{m=1}^\infty B_m(\mu)C^{3/2}_m(2x-1)\biggr\}, \eeq
where $f_S(\mu)$ and $\bar f_S(\mu)$, $B_m(\mu)$, and
$C_m^{3/2}(t)$ are the vector and scalar decay constants,
Gegenbauer moments, and Gegenbauer  polynomials for the scalars,
respectively.

Because of the charge conjugation invariance, neutral scalar mesons
cannot be produced by the vector current and thus
 \begin{equation}
 f_{\sigma}=f_{f_0}=f_{a_0^0}=0.
 \end{equation}
For other scalar mesons, there exists a relation between
the vector and scalar decay constants,
\beq
 \bar f_S &=& \mu_S f_S \;\;\;\;  {\rm and} \;\;\;\;  \mu_S = \frac{m_S}{m_2(\mu)-m_1(\mu)}\;,
\eeq
where $m_1$ and $m_2$ are the
running current quark masses in the scalars. For the neutral scalar mesons
$f_0$, $a_0^0$ and $\sigma$, $f_S$ vanishes, but the quantity
$\bar f_S=f_S\mu_S$ remains finite.

The values for scalar decay constants and Gegenbauer moments in the scalar meson distribution amplitudes have been investigated at scale $\mu=1~
\mbox{GeV}$ in Ref.~\cite{Cheng06:B2SP}:
\beq
\bar f_{a_0}&=& 0.365 \pm 0.020~{\rm GeV}, \quad B_1= -0.93 \pm 0.10, \quad  B_3= 0.14 \pm 0.08\;\;\; \rm{(S1)} \;,  \non
\bar f_{\kappa}&=& 0.340 \pm 0.020~{\rm GeV}, \quad B_1= -0.92 \pm 0.11, \quad  B_3= 0.15 \pm 0.09\;\;\; \rm{(S1)} \;, \\
\bar f_{f_0}&=& 0.370 \pm 0.020~{\rm GeV}, \quad B_1= -0.92 \pm 0.11, \quad  B_3= 0.15 \pm 0.09\;\;\; \rm{(S1)}\;; \nonumber
\eeq
\beq
\bar f_{a_0(1450)}&=& -0.280 \pm 0.030~{\rm GeV}, \quad B_1= 0.89 \pm 0.20, \quad  B_3= -1.38 \pm 0.18\;\;\; \rm{(S1)}\;,   \non
\bar f_{a_0(1450)}&=&  0.460 \pm 0.050~{\rm GeV}, \quad B_1= -0.58 \pm 0.12, \quad  B_3= -0.49 \pm 0.15\;\;\; \rm{(S2)}\;;
\eeq
\beq
\bar f_{K_0^*(1430)}&=& -0.300 \pm 0.030~{\rm GeV}, \quad B_1= 0.58 \pm 0.07, \quad  B_3= -1.20 \pm 0.08\;\;\; \rm{(S1)}\;,  \non
\bar f_{K_0^*(1430)}&=& 0.445 \pm 0.050~{\rm GeV}, \quad B_1= -0.57 \pm 0.13, \quad  B_3= -0.42 \pm 0.22\;\;\; \rm{(S2)}\;;
\eeq
\beq
\bar f_{f_0(1500)}&=& -0.255 \pm 0.030~{\rm GeV}, \quad B_1= 0.80 \pm 0.40, \quad  B_3= -1.32 \pm 0.14\;\;\; \rm{(S1)}\;,  \non
\bar f_{f_0(1500)}&=& 0.490 \pm 0.050~{\rm GeV}, \quad B_1= -0.48 \pm 0.11, \quad  B_3= -0.37 \pm 0.20\;\;\; \rm{(S2)}\;.
\eeq

As for the twist-3 distribution amplitudes $\phi_{S}^S$ and
$\phi_{S}^T$, we adopt the asymptotic forms:
\beq
\phi^S_{S}&=& \frac{1}{2\sqrt {2N_c}}\bar f_{S},\,\,\,\,\,\,\,\phi_{S}^T=
\frac{1}{2\sqrt {2N_c}}\bar f_{S}(1-2x).
\eeq
Note that for the distribution amplitudes of strange scalar meson,
$x$ stands for the momentum fraction carrying by $s$ quark.

For 
pseudoscalar meson($P$), the wave function can be generally
defined as,
\beq
\Phi_P(x) &=& \frac{i}{\sqrt{2 N_c}} \gamma_5
\left\{\psl \phi_P^A(x)+ m_0^P \phi_P^P(x) + m_0^P (\nsl \vsl -1)
\phi_P^T(x)\right\}_{\alpha\beta}\;,
\eeq
where $\phi_P^{A,P,T}$ and $m_0^P$ are the distribution amplitudes and chiral scale parameter
of the pseudoscalar meson, respectively.

For the wave functions of vector meson($V$), one
longitudinal($L$) polarization is involved, and
 can be written as,
\beq
\Phi^L_V(x)&=& \frac{1}{\sqrt{2 N_c}} \left\{
m_V \epsl_V^{*L} \phi_V(x) + \epsl^{*L}_V \psl \phi_V^t(x)+ m_V
\phi_V^s(x)\right\}_{\alpha\beta} \;,
\eeq
where $\epsilon_V^{L}$ denotes the longitudinal polarization
vector of vector mesons, satisfying $P \cdot \epsilon=0$, $\phi_V$
and $\phi_V^{t,s}$, and $m_V$ are the leading twist and twist-3 distribution amplitudes, and mass
of the vector meson, respectively.
For the distribution
amplitudes of pseudoscalar $\phi_P^{A,P,T}$, and longitudinal polarization, $\phi_V$ and $\phi_V^{t,s}$ to be used
in this work, we will adopt the same forms as that in the literatures
(See Ref.~\cite{Xiao1:bcdecay} and references therein).

\section{perturbative calculations in the pQCD approach} \label{sec:2}

\begin{figure}[t,b]
\vspace{-0.5cm} \centerline{\epsfxsize=16 cm \epsffile{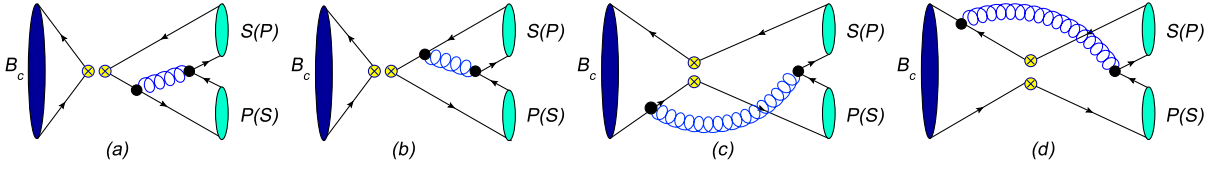}}
\vspace{0.2cm} \caption{ Typical Feynman diagrams for two-body
charmless hadronic $B_c \to SP(PS)$ decays at leading order. By replacing the pseudoscalar meson $P$ with
the vector meson $V$, which will lead to the diagrams for $B_c \to SV(VS)$ modes. }
 \label{fig:fig1}
\end{figure}

From the effective Hamiltonian~(\ref{heff}), there are 4 types of
diagrams contributing to the $B_c \to M_2 M_3$ decays as
illustrated in Fig.~\ref{fig:fig1}, which result in the Feynman
decay amplitudes ${\cal F}_{fa}^{M_2 M_3}$ and ${\cal M}_{na}^{M_2 M_3}$,  where
the subscripts ${fa}$ and ${na}$  are the abbreviations of
factorizable and non-factorizable annihilation contributions, respectively.
Operators $O_{1,2}$ are $(V-A)(V-A)$ currents, we
therefore can combine all contributions from these diagrams and
obtain the total decay amplitude as,
\beq
{\cal A}(B_c \to  M_2
M_3) &=& V_{cb}^* V_{uD} \left\{f_{B_c} {\cal F}_{fa}^{M_2 M_3}a_1 +
{\cal M}_{na}^{M_2 M_3} C_1 \right\} \; ,
\label{eq:amt}
\eeq
where $a_1=C_1/3+C_2$.  In the next two subsections we will give the
explicit expressions of ${\cal F}_{fa}^{M_2M_3}$, ${\cal M}_{na}^{ M_2 M_3}$ and
the decay amplitude ${\cal A}(B_c \to  M_2 M_3)$ for $B_c\to M_2 M_3$
decays: including 32 $B_c \to SP(PS)$ and 30 $B_c \to SV(VS)$ decay modes.

\subsection{  $B_c \to SP(PS)$ decays}

In this subsection, we will present the factorization formulas for
charmless hadronic $B_c \to SP(PS)$ decays. From the first two
diagrams of Fig.~\ref{fig:fig1}, i.e., (a) and (b), by perturbative
QCD calculations, we obtain the decay amplitude
for factorizable annihilation contributions as follows,
\beq
{\cal F}_{fa}^{SP} &=& 8 \pi C_F m_{B_c}^2 \int_0^1 d x_{2} dx_{3}\,
\int_{0}^{\infty} b_2 db_2 b_3 db_3\, \non && \times
\left\{h_{fa}(1-x_{3},x_{2},b_{3},b_{2})E_{fa}(t_{a}) \left[x_{2}
\phi_{S}(x_2)\phi^A_{P}(x_3)+2 r_S r_0^P\phi_{P}^P(x_3)
\right.\right. \non && \left.\left. \times
 \left((x_2 + 1)\phi^{S}_{S}(x_2)+ (x_2 -
1)\phi^{T}_{S}(x_{2})\right)\right]+h_{fa}(x_{2},1-x_{3},b_{2},b_{3})E_{fa}(t_{b})\right.
\non && \left.
 \times\left[ (x_3
-1) \phi_{S}(x_2) \phi_{P}^A(x_3) + 2 r_S r_0^P
\phi_{S}^S(x_2) \left( (x_3 -2)\phi_{P}^P(x_3)- x_3
\phi_{P}^T(x_3)\right)\right] \right\}\;,
\label{eq:ab}
\eeq
where $\phi_{S(P)}$ corresponds to the distribution amplitudes of
mesons $S(P)$, $r_{S}= m_{S}/m_{B_c}$, $r_0^{P}= m_{0}^{P}/m_{B_c}$, and
$C_F=4/3$ is a color factor.
The
function $h_{fa}$, the scales $t_i$ and $E_{fa}(t)$ can be found
in Appendix~B of Ref.~\cite{Xiao1:bcdecay}.

For the nonfactorizable diagrams (c) and (d) in Fig.~\ref{fig:fig1}, all three meson wave
functions are involved. The integration of $b_3$ can be performed
using $\delta$ function $\delta(b_3-b_2)$, leaving only integration
of $b_1$ and $b_2$. The corresponding decay amplitude is
\beq
 {\cal M}_{na}^{SP} &=& \frac{16 \sqrt{6}}{3}\pi C_F m_{B_c}^2
\int_{0}^{1}d x_{2}\,d x_{3}\,\int_{0}^{\infty} b_1d b_1 b_2d b_2\,
 \non && \times \left\{h_{na}^{c}(x_2,x_3,b_1,b_2) E_{na}(t_c)
\left[(r_c - x_3 +1) \phi_{S}(x_2)\phi_{P}^A (x_{3})+ r_S
r_0^P\left(\phi_{S}^S(x_2) \right.\right. \right.\non && \left. \left.
\left. \times ((3 r_c + x_2 -x_3 +1) \phi_{P}^P(x_3)-(r_c -x_2 -x_3
+1)\phi_{P}^T(x_3))+\phi_{S}^T(x_2)\right.\right.\right.
\non &&
\left.\left. \left. \times ((r_c-x_2 -x_3 +1) \phi_{P}^P(x_3)+(r_c
-x_2 +x_3 -1)\phi_{P}^T(x_3))\right)\right]-E_{na}(t_d) \right.
\non
&& \left.
\times\left[ (r_b + r_c +x_2 -1) \phi_{S}(x_2) \phi
_{P}^A(x_3) + r_S r_0^P \left(\phi_{S}^S(x_2)((4 r_b
 +r_c +x_2 -x_3\right.\right.\right.\non
&& \left.
\left.\left. -1)\phi_{P}^P(x_3)-(r_c + x_2 +x_3 -1)\phi_{P}^T(x_3))
+\phi_{S}^T(x_2)((r_c + x_2 +x_3 -1)\right.\right.\right. \non
&&
\left.\left.\left. \times\phi_{P}^P(x_3)-(r_c +x_2 -x_3 -1)
\phi_{P}^T(x_3))\right)\right]
h_{na}^{d}(x_2,x_3,b_1,b_2)\right\}\;,\label{eq:cd}
\eeq
where $r_{b}= m_{b}/m_{B_c}$, $r_{c}= m_{c}/m_{B_c}$, and $r_b+r_c \approx 1$ in $B_c$ meson.

Likewise, we can get the analytic factorization formulas of the contributions from $B_c \to PS$ decays easily,
\beq
{\cal F}_{fa}^{PS} &=& 8 \pi C_F m_{B_c}^2 \int_0^1 d x_{2} dx_{3}\,\int_{0}^{\infty} b_2 db_2 b_3 db_3\,
\non && \times
\left\{h_{fa}(1-x_{3},x_{2},b_{3},b_{2})E_{fa}(t_{a}) \left[x_{2}\phi_{P}^A(x_2)\phi_{S}(x_3)-2 r_0^P r_S\phi_{S}^S(x_3)
\right.\right. \non && \left.\left. \times
\left((x_2 + 1)\phi^{P}_{P}(x_2)+ (x_2 -1)\phi^{T}_{P}(x_{2})\right)\right]+h_{fa}(x_{2},1-x_{3},b_{2},b_{3})E_{fa}(t_{b})
\right.\non && \left. \times
\left[ (x_3 -1) \phi_{P}^A(x_2) \phi_{S}(x_3) - 2 r_0^P r_S \phi_{P}^P(x_2) \left( (x_3 -2)\phi_{S}^S(x_3)- x_3
\phi_{S}^T(x_3)\right)\right] \right\}\;,
\label{eq:ab'}
 \eeq
\beq
{\cal M}_{na}^{PS} &=& \frac{16 \sqrt{6}}{3}\pi C_F m_{B_c}^2\int_{0}^{1}d x_{2}\,d x_{3}\,\int_{0}^{\infty} b_1d b_1 b_2d b_2\,
 \non && \times
 \left\{h_{na}^{c}(x_2,x_3,b_1,b_2) E_{na}(t_c) \left[(r_c - x_3 +1) \phi_{P}^A(x_2)\phi_{S} (x_{3})- r_0^P r_S\left(\phi_{P}^P(x_2)
  \right.\right. \right.\non && \left. \left. \left. \times
((3 r_c + x_2 -x_3 +1) \phi_{S}^S(x_3)-(r_c -x_2 -x_3 +1)\phi_{S}^T(x_3))+\phi_{P}^T(x_2)
\right.\right.\right. \non && \left.\left. \left. \times
((r_c-x_2 -x_3 +1) \phi_{S}^S(x_3)+(r_c -x_2 +x_3 -1)\phi_{P}^T(x_3))\right)\right]-E_{na}(t_d)
\right. \non && \left.\times
\left[ (r_b + r_c +x_2 -1) \phi_{P}^A(x_2) \phi_{S}(x_3) - r_0^P r_S \left(\phi_{P}^P(x_2)((4 r_b +r_c +x_2 -x_3
\right.\right.\right.\non && \left. \left.\left.
-1)\phi_{S}^S(x_3)-(r_c + x_2 +x_3 -1)\phi_{S}^T(x_3)) +\phi_{P}^T(x_2)((r_c + x_2 +x_3 -1)
\right.\right.\right. \non && \left.\left.\left. \times
\phi_{S}^S(x_3)-(r_c +x_2 -x_3 -1) \phi_{S}^T(x_3))\right)\right] h_{na}^{d}(x_2,x_3,b_1,b_2)\right\}\;.
\label{eq:cd'}
\eeq

Based on Eqs.(\ref{eq:amt}-\ref{eq:cd'}),
we can write down the total decay amplitudes for 32 $B_c \to SP(PS)$
decays straightforwardly,
\beq
{\cal A}(B_c \to a^+ \pi^0) &=& V_{cb}^* V_{ud} \left\{[f_{B_c} {\cal F}_{fa}^{a \pi}a_1 + {\cal M}_{na}^{a \pi} C_1 ] \right. \non && \left. -[f_{B_c}
{\cal F}_{fa}^{\pi a} a_1 + {\cal M}_{na}^{\pi a } C_1 ]  \right\}/\sqrt{2}\;, \label{eq:a0pi0}  \\
{\cal A}(B_c \to a^0 \pi^+) &=& V_{cb}^* V_{ud} \left\{[f_{B_c} {\cal F}_{fa}^{\pi a}a_1 + {\cal M}_{na}^{\pi a} C_1 ] \right. \non && \left. -[f_{B_c}
{\cal F}_{fa}^{a \pi} a_1 + {\cal M}_{na}^{a \pi} C_1 ]  \right\}/\sqrt{2}\;;\label{eq:pipa0}
\eeq
\beq
{\cal A}(B_c \to a^+ \eta) &=& V_{cb}^* V_{ud} \left\{[f_{B_c} {\cal F}_{fa}^{a \eta_{q}}a_1 + {\cal M}_{na}^{a \eta_{q}} C_1 ] \right. \non && \left.
+ [f_{B_c} {\cal F}_{fa}^{\eta_{q} a }a_1 + {\cal M}_{na}^{\eta_{q} a } C_1 ]\right\}\cos\phi\;, \label{eq:a0peta}\\
{\cal A}(B_c \to a^+ \eta') &=& V_{cb}^* V_{ud} \left\{[f_{B_c} {\cal F}_{fa}^{a \eta_{q}}a_1 + {\cal M}_{na}^{a \eta_{q}} C_1] \right. \non && \left.
+ [f_{B_c} {\cal F}_{fa}^{\eta_{q} a}a_1 + {\cal M}_{na}^{\eta_{q} a} C_1] \right\}\sin\phi\;;\label{eq:a0peta'}
\eeq
\beq
{\cal A}(B_c \to f \pi^+) &=& V_{cb}^* V_{ud} \left\{[f_{B_c} {\cal F}_{fa}^{\pi f_0^{q}}a_1 + {\cal M}_{na}^{\pi f_0^{q}} C_1 ] \right. \non && \left.
+ [f_{B_c} {\cal F}_{fa}^{f_0^{q} \pi }a_1 + {\cal M}_{na}^{f_0^{q} \pi } C_1 ]\right\}\cos\theta_0\;,\label{eq:pipf0}\\
{\cal A}(B_c \to f' \pi^+) &=& V_{cb}^* V_{ud} \left\{[f_{B_c} {\cal F}_{fa}^{\pi f_0^{q}}a_1 + {\cal M}_{na}^{\pi f_0^{q}} C_1 ] \right. \non && \left.
+ [f_{B_c} {\cal F}_{fa}^{f_0^{q} \pi }a_1 + {\cal M}_{na}^{f_0^{q} \pi } C_1 ]\right\}\sin\theta_0\;;\label{eq:pipf0'}
\eeq
\beq
{\cal A}(B_c \to K_0^{*+} \overline{K}^0) &=& V_{cb}^* V_{ud} \left\{f_{B_c} {\cal F}_{fa}^{K_0^{*} \overline{K}}a_1 +
{\cal M}_{na}^{K_0^{*} \overline{K}} C_1 \right\}\;,\label{eq:kbk0p}\\
{\cal A}(B_c \to \overline{K}_0^{*0} K^{+}) &=& V_{cb}^* V_{ud} \left\{f_{B_c} {\cal F}_{fa}^{\overline{K}_0^{*} K^{+}}a_1 +
{\cal M}_{na}^{\overline{K}_0^{*} K^{+}} C_1 \right\}\;;\label{eq:k0sbkp}
\eeq
\beq
{\cal A}(B_c \to  K_0^{*0} \pi^+) &=& V_{cb}^* V_{us} \left\{f_{B_c} {\cal F}_{fa}^{K_0^{*} \pi } a_1 + {\cal M}_{na}^{ K_0^{*} \pi} C_1 \right\}  \;,\label{eq:k0pip} \\
&=& \sqrt{2} {\cal A}(B_c \to  K_0^{*+} \pi^0) \label{eq:k0ppi0}\;;
\eeq
\beq
{\cal A}(B_c \to a^+  K^0) &=& V_{cb}^* V_{us} \left\{f_{B_c} {\cal F}_{fa}^{K a} a_1 + {\cal M}_{na}^{K a} C_1 \right\} \;,\label{eq:k0a0p}  \\
&=& \sqrt{2} {\cal A}(B_c \to  K^+ a^0)\label{eq:kpa0}\;;
\eeq
\beq
{\cal A}(B_c \to K_0^{*+} \eta) &=& V_{cb}^* V_{us} \left\{f_{B_c} \left [{\cal F}_{fa}^{K_0^* \eta_q} \cos\phi - {\cal F}_{fa}^{\eta_s K_0^* } \sin\phi \right ] a_1\right.
\non && \left.  + \left [{\cal M}_{na}^{K_0^* \eta_q} \cos\phi - {\cal M}_{na}^{\eta_s K_0^* } \sin\phi \right ] C_1 \right\}\;,\label{eq:k0eta}\\
{\cal A}(B_c \to K_0^{*+} \eta') &=&  V_{cb}^* V_{us} \left\{f_{B_c} \left [{\cal F}_{fa}^{K_0^* \eta_q} \sin\phi + {\cal F}_{fa}^{\eta_s K_0^* } \cos\phi \right ] a_1\right.
\non && \left.  + \left [{\cal M}_{na}^{K_0^* \eta_q} \sin\phi + {\cal M}_{na}^{\eta_s K_0^* } \cos\phi\right ] C_1 \right\}\;; \label{eq:k0etap}
\eeq
\beq
{\cal A}(B_c \to f K^+) &=& V_{cb}^* V_{us} \left\{f_{B_c} \left [{\cal F}_{fa}^{K  f_0^q} \cos\theta_0 - {\cal F}_{fa}^{f_0^s K } \sin\theta_0 \right ] a_1\right.
\non && \left.  + \left [{\cal M}_{na}^{K f_0^q} \cos\theta_0 - {\cal M}_{na}^{f_0^s K } \sin\theta_0 \right ] C_1 \right\}\;,\label{eq:kpf0}\\
{\cal A}(B_c \to f' K^+) &=&  V_{cb}^* V_{us} \left\{f_{B_c} \left [{\cal F}_{fa}^{K f_0^q} \sin\theta_0 + {\cal F}_{fa}^{f_0^s K } \cos\theta_0 \right ] a_1\right.
\non && \left.  + \left [{\cal M}_{na}^{K f_0^q} \sin\theta_0 + {\cal M}_{na}^{f_0^s K } \cos\theta_0\right ] C_1 \right\}\;. \label{eq:kpf0'}
\eeq

\subsection{  $B_c \to SV(VS)$ decays}

After the replacement of the pseudoscalar meson $P$ with
the vector meson $V$ in Figure~\ref{fig:fig1}, we will get the Feynman diagrams for pure annihilation $B_c \to SV(VS)$ modes at leading order.
By following the same procedure as stated in the above subsection,
we can obtain the analytic decay amplitudes for $B_c \to SV$ decays,
\beq
{\cal F}_{fa}^{SV} &=&-8 \pi C_F m_{B_c}^2 \int_0^1 d x_{2} dx_{3}\, \int_{0}^{\infty} b_2 db_2 b_3 db_3\,
\non && \times
\left\{h_{fa}(1-x_{3},x_{2},b_{3},b_{2})E_{fa}(t_{a}) \left[x_{2} \phi_{S}(x_2)\phi_{V}(x_3)-2 r_S r_V\phi_{V}^s(x_3)
\right.\right. \non && \left.\left. \times
\left((x_2 + 1)\phi^{S}_{S}(x_2)+ (x_2 -1)\phi^{T}_{S}(x_{2})\right)\right]+ h_{fa}(x_{2},1-x_{3},b_{2},b_{3})E_{fa}(t_{b})
\right. \non && \left. \times
\left[ (x_3 -1) \phi_{S}(x_2) \phi_{V}(x_3) - 2 r_S r_V \phi_{S}^S(x_2) \left( (x_3 -2)\phi_{V}^s(x_3)- x_3 \phi_{V}^t(x_3)\right)\right]\right\}\;,
\label{eq:ab1}\\
{\cal M}_{na}^{SV} &=& -\frac{16 \sqrt{6}}{3}\pi C_F m_{B_c}^2 \int_{0}^{1}d x_{2}\,d x_{3}\,\int_{0}^{\infty} b_1d b_1 b_2d b_2\,
 \non && \times
 \left\{h_{na}^{c}(x_2,x_3,b_1,b_2) E_{na}(t_c) \left[(r_c - x_3 +1) \phi_{S}(x_2)\phi_{V}(x_{3})- r_S r_V\left(\phi_{S}^S(x_2)
 \right.\right. \right.\non && \left. \left. \left. \times
 ((3 r_c + x_2 -x_3 +1) \phi_{V}^s(x_3)-(r_c -x_2 -x_3 +1)\phi_{V}^t(x_3))+\phi_{S}^T(x_2)
 \right.\right.\right. \non && \left.\left. \left. \times
 ((r_c-x_2 -x_3 +1) \phi_{V}^s(x_3)+(r_c -x_2 +x_3 -1)\phi_{V}^t(x_3))\right)\right]-E_{na}(t_d)
 \right. \non && \left. \times
 \left[ (r_b + r_c +x_2 -1) \phi_{S}(x_2) \phi_{V}(x_3) - r_S r_V \left(\phi_{S}^S(x_2)((4 r_b +r_c +x_2 -x_3
 \right.\right.\right.\non && \left. \left.\left.
 -1)\phi_{V}^s(x_3)-(r_c + x_2 +x_3 -1)\phi_{V}^t(x_3)) +\phi_{S}^T(x_2)((r_c + x_2 +x_3 -1)
 \right.\right.\right. \non && \left.\left.\left. \times
 \phi_{V}^s(x_3)-(r_c +x_2 -x_3 -1) \phi_{V}^t(x_3))\right)\right] h_{na}^{d}(x_2,x_3,b_1,b_2)\right\}\;,
 \label{eq:cd1}
 \eeq
with $r_V=m_V/m_{B_c}$.

Similarly, the factorization formulas for $B_c \to V S$ decays can be easily
obtained but with the simple replacements in Eqs.~(\ref{eq:ab1},\ref{eq:cd1}) as follows,
\beq
 \phi_S &\longleftrightarrow &  \phi_V, \quad  \phi_S^S \longleftrightarrow   \phi_V^s,
\quad \phi_S^T \longleftrightarrow   \phi_V^t,
\quad r_S \longleftrightarrow   r_V \;. \label{eq:cd2}
\eeq

The total decay amplitudes of the 30 $B_c \to SV(VS)$ decays can therefore
be written as,
\beq
{\cal A}(B_c \to a^+ \rho^0) &=& V_{cb}^* V_{ud} \left\{[f_{B_c} {\cal F}_{fa}^{a \rho}a_1 + {\cal M}_{na}^{a \rho} C_1 ] \right. \non && \left. -[f_{B_c}
{\cal F}_{fa}^{\rho a} a_1 + {\cal M}_{na}^{\rho a } C_1 ]  \right\}/\sqrt{2}\;, \label{eq:a0rho0}  \\
{\cal A}(B_c \to a^0 \rho^+) &=& V_{cb}^* V_{ud} \left\{[f_{B_c} {\cal F}_{fa}^{\rho a}a_1 + {\cal M}_{na}^{\rho a} C_1 ] \right. \non && \left. -[f_{B_c}
{\cal F}_{fa}^{a \rho} a_1 + {\cal M}_{na}^{a \rho} C_1 ]  \right\}/\sqrt{2}\;;\label{eq:rhopa0}
\eeq
\beq
{\cal A}(B_c \to a^+ \omega) &=& V_{cb}^* V_{ud} \left\{[f_{B_c} {\cal F}_{fa}^{a \omega}a_1 + {\cal M}_{na}^{a \omega} C_1 ] \right. \non && \left.
+ [f_{B_c} {\cal F}_{fa}^{\omega a }a_1 + {\cal M}_{na}^{\omega a } C_1 ]\right\} /\sqrt{2}\;;\label{eq:a0pomega}
\eeq
\beq
{\cal A}(B_c \to f \rho^+) &=& V_{cb}^* V_{ud} \left\{[f_{B_c} {\cal F}_{fa}^{\rho f_0^{q}}a_1 + {\cal M}_{na}^{\rho f_0^{q}} C_1 ] \right. \non && \left.
+ [f_{B_c} {\cal F}_{fa}^{f_0^{q} \rho }a_1 + {\cal M}_{na}^{f_0^{q} \rho } C_1 ]\right\}\cos\theta_0\;,\label{eq:rhopf0}\\
{\cal A}(B_c \to f' \rho^+) &=& V_{cb}^* V_{ud} \left\{[f_{B_c} {\cal F}_{fa}^{\rho f_0^{q}}a_1 + {\cal M}_{na}^{\rho f_0^{q}} C_1 ] \right. \non && \left.
+ [f_{B_c} {\cal F}_{fa}^{f_0^{q} \rho }a_1 + {\cal M}_{na}^{f_0^{q} \rho } C_1 ]\right\}\sin\theta_0\;;\label{eq:rhopf0'}
\eeq
\beq
{\cal A}(B_c \to K_0^{*+} \overline{K}^{*0}) &=& V_{cb}^* V_{ud} \left\{f_{B_c} {\cal F}_{fa}^{\overline{K}^* K_0^{*}}a_1 +
{\cal M}_{na}^{\overline{K}^* K_0^{*}} C_1 \right\}\;,\label{eq:ksbk0p}\\
{\cal A}(B_c \to \overline{K}_0^{*0} K^{*+}) &=& V_{cb}^* V_{ud} \left\{f_{B_c} {\cal F}_{fa}^{\overline{K}_0^{*} K^{*+}}a_1 +
{\cal M}_{na}^{\overline{K}_0^{*} K^{*+}} C_1 \right\}\;;\label{eq:k0sbksp}
\eeq
\beq
{\cal A}(B_c \to  K_0^{*0} \rho^+) &=& V_{cb}^* V_{us} \left\{f_{B_c} {\cal F}_{fa}^{K_0^{*} \rho } a_1 + {\cal M}_{na}^{ K_0^{*} \rho} C_1 \right\}  \;,\label{eq:k0rhop} \\
&=& \sqrt{2} {\cal A}(B_c \to  K_0^{*+} \rho^0) \label{eq:k0prho0}\;;
\eeq
\beq
{\cal A}(B_c \to a^+  K^{*0}) &=& V_{cb}^* V_{us} \left\{f_{B_c} {\cal F}_{fa}^{K^* a} a_1 + {\cal M}_{na}^{K^* a} C_1 \right\} \;,\label{eq:ks0a0p}  \\
&=& \sqrt{2} {\cal A}(B_c \to  K^{*+} a^0)\label{eq:kspa0}\;;
\eeq
\beq
{\cal A}(B_c \to K_0^{*+} \omega) &=& V_{cb}^* V_{us} \left\{f_{B_c} {\cal F}_{fa}^{K_0^* \omega} a_1 + {\cal M}_{na}^{K_0^* \omega} C_1 \right\}/\sqrt{2}\;,\label{eq:k0omega}\\
{\cal A}(B_c \to K_0^{*+} \phi) &=& V_{cb}^* V_{us} \left\{f_{B_c} {\cal F}_{fa}^{\phi K_0^* } a_1 + {\cal M}_{na}^{\phi K_0^* } C_1 \right\}\;; \label{eq:phik0p}
\eeq
\beq
{\cal A}(B_c \to f K^{*+}) &=& V_{cb}^* V_{us} \left\{f_{B_c} \left [{\cal F}_{fa}^{K^*  f_0^q} \cos\theta_0 - {\cal F}_{fa}^{f_0^s K^* } \sin\theta_0 \right ] a_1\right.
\non && \left.  + \left [{\cal M}_{na}^{K^* f_0^q} \cos\theta_0 - {\cal M}_{na}^{f_0^s K^* } \sin\theta_0 \right ] C_1 \right\}\;,\label{eq:kspf0}\\
{\cal A}(B_c \to f' K^{*+}) &=&  V_{cb}^* V_{us} \left\{f_{B_c} \left [{\cal F}_{fa}^{K^* f_0^q} \sin\theta_0 + {\cal F}_{fa}^{f_0^s K^* } \cos\theta_0 \right ] a_1\right.
\non && \left.  + \left [{\cal M}_{na}^{K^* f_0^q} \sin\theta_0 + {\cal M}_{na}^{f_0^s K^* } \cos\theta_0\right ] C_1 \right\}\;. \label{eq:kspf0'}
\eeq


\section{Numerical Results and Discussions}\label{sec:3}

In this section, we will make the theoretical predictions on
the {\it CP}-averaged
BRs for those considered
$B_c \to SP, SV$ decay modes. First of all, the central values of the input
parameters to be used are given in the following,

\begin{itemize}
\item {Masses (GeV):}
\beq
 m_W &=& 80.41\;,\;\;\;\;\;\;  \quad m_{B_c} = 6.286\;,\;\;\;\;\;\;\;  \quad m_b = 4.8\;,\;\;\;\;\;\;\;\;\;\; \quad m_c = 1.5\;; \non
m_\phi &=& 1.02\;,\;\;\;\;\;\;\;  \quad  m_{K^*}= 0.892\;,\;\;\;\;\;\;\; \quad  m_{\rho} = 0.770\;,\;\;\;\;\;\;\;\;\;\;  m_{\omega}=0.782\;;   \non
 m_{a_0} &=& 0.985\;,\;\;\;\;\;\;\;  \quad  m_{\kappa} = 0.800\;,\;\;\;\;\;\;\ \quad   m_{\sigma}= 0.600\;,\;\;\;\;\;\;  \quad  m_{f_0} = 0.980\;;    \non
m_{a_0(1450)} &=& 1.474\;,  \quad  m_{{K_0^*}(1430)} =1.425\;, \quad  m_{f_0(1370)}= 1.350\;,  \quad  m_{f_0(1500)} = 1.505\;;  \non
m_{0}^\pi &=& 1.4\;,\;\;\;\;\;\;\;\;\;\;\  \quad  m_{0}^K = 1.6\;,\;\;\;\;\;\;\;\;\ \quad  m_{0}^{\eta_q}= 1.08\;,\;\;\;\;\;\;\;  \quad  m_{0}^{\eta_s} = 1.92\;. \label{eq:mass}
\eeq
\item {Decay constants (GeV):}
\beq
f_{\phi} &=& 0.231\;,\quad  f_{\phi}^T = 0.200\;,  \quad f_{K^*} = 0.217\;, \quad f_{K^*}^T = 0.185\;; \non
f_{\rho}&=& 0.209\;, \quad f^T_{\rho}= 0.165\;,\;\; \quad f_{\omega}= 0.195\;,\;\; \quad f_{\omega}^T = 0.145\;; \non
f_{\pi} &=& 0.131\;, \quad f_{K} = 0.16\;,\;\;  \quad f_{B_c} = 0.489\;. \label{eq:dconst}
 \eeq
\item { QCD scale and $B_c$ meson lifetime:}
\beq
\Lambda_{\overline{\mathrm{MS}}}^{(f=4)} &=& 0.250\; {\rm GeV},  \quad
\tau_{B_c}= 0.46\; {\rm ps}.
\eeq

\end{itemize}

Here, we adopt the Wolfenstein
parametrization, and the updated parameters $A=0.814$,
 $\lambda=0.2257$, $\bar{\rho}=0.135$, and $\bar{\eta}=0.349$~\cite{Amsler08:pdg} for the CKM matrix.
 In numerical calculations, central values of the input parameters will be
used implicitly unless otherwise stated.


For $B_c \to SP, SV$ decays, the decay rate can be written as
\beq
\Gamma =\frac{G_{F}^{2}m^{3}_{B_c}}{32 \pi  } (1-r_S^2) |{\cal A}(B_c
\to M_2 M_3)|^2\;,
\eeq
where
 the corresponding decay amplitudes ${\cal A}$ have been
given explicitly in Eqs.~(\ref{eq:a0pi0}-\ref{eq:kpf0'}) and
Eqs.~(\ref{eq:a0rho0}-\ref{eq:kspf0'}). Using the decay amplitudes
obtained in last section, it is straightforward to calculate the
{\it CP}-averaged BRs with uncertainties as presented in
Tables~\ref{tab:bcsp-1s0}-\ref{tab:bcsv-2s1}. The dominant errors come from the uncertainties of charm quark mass
$m_c=1.5 \pm 0.15$ GeV, the scalar decay constants $\bar f_S$,
 the Gegenbauer moments $a_i$ of the relevant pseudoscalar or vector
meson distribution amplitudes, the Gegenbauer moments $B_i$ of the scalar meson distribution amplitudes,
 and the chiral enhancement factors
$m_0^{\pi}=1.4 \pm 0.3$ GeV  and $m_0^{K}=1.6 \pm 0.1$ GeV, respectively.

Among the considered $B_c\to SP, SV$ decays, the
pQCD predictions for the {\it CP}-averaged BRs of those
$\Delta S = 0$ processes are basically much larger than those of $\Delta S =1$
channels (one of the two final state mesons is a strange one),
the main reason is the enhancement of  the large CKM factor $|V_{ud}/V_{us}|^2 \sim 19$
for those $\Delta S = 0$ decays as generally expected.  Maybe there exist no such large differences for
certain decays, which is just because the enhancement arising from the
CKM factor is partially cancelled by the difference between the
magnitude of individual decay amplitude.
The pQCD predictions for the {\it CP}-averaged BRs of considered
$B_c$ decays vary in the range of $10^{-5}$ to  $10^{-8}$.
For $B_c \to a_0(1450)^+ \pi^0$ decay with rate of $10^{-5} \sim 10^{-6}$ for example, 
 we show the decay amplitudes arising from both factorization
and nonfactorization annihilation contributions explicitly(in unit of $10^{-3}$ ${\rm GeV}^3$),
 \beq
{\cal A}_{fa}(B_c \to a_0(1450)^+ \pi^0) &=& 0.292 + {\it i} 2.489; \quad {\cal A}_{na}(B_c \to a_0(1450)^+ \pi^0) =  6.717 + {\it i} 7.508;
 \eeq
in S1, while
 \beq
{\cal A}_{fa}(B_c \to a_0(1450)^+ \pi^0) &=& 0.553 - {\it i} 0.356; \quad {\cal A}_{na}(B_c \to a_0(1450)^+ \pi^0) =  3.161 - {\it i} 5.137.
 \eeq
in S2, where the central values are quoted for clarification. One can find that
the dominant nonfactorizable decay amplitude governs this channel and subsequently results in the large branching ratio in both scenarios, which can be
seen in Table~\ref{tab:bcsp-2s0}. The
 other modes with large decay rates can be
analyzed similarly.

As discussed in Ref.~\cite{ekou09:ncbc}, the $B_c$ decays with the branching ratio of $10^{-6}$ can be measured at
the LHC. Hence our pQCD predicted BRs with $10^{-6}$ or larger for these $B_c \to SP, SV$ decays
are expected to be measured in the ongoing LHCb experiments, which will be very helpful to study the physical contents
of the scalars and the involved QCD dynamics and annihilation mechanism in the these considered channels.
Moreover,
there is no CP violation for all these decays within the SM,
since there
is only one kind of tree operator involved in the decay amplitude of
all considered $B_c$ decays, which can be seen from Eq.~(\ref{eq:amt}).

\subsection{ $B_c \to a_0 (P, V)$ and $B_c \to a_0(1450) (P, V)$ decays}

\begin{table}[t]
\caption{The pQCD predictions of branching ratios(BRs) for the $\Delta S =0$ processes
of charmless hadronic $B_c \to
(a_0, \kappa, \sigma, f_0)(\pi, K, \eta, \eta')$ decays in S1. The source of the
dominant errors is explained in the text.}
\label{tab:bcsp-1s0}
\begin{center}\vspace{-0.6cm}
\begin{tabular}[t]{l|l}
 \hline \hline
 $\Delta S =0$     &                                              \\
 Decay modes       &    BRs $(10^{-6})$         \\
\hline
$\rm{B_c \to a_0^+ \pi^0}$ & $6.5^{+2.3}_{-1.5}(m_c)^{+0.9}_{-0.6}(\bar f_S)^{+2.1}_{-1.4}(a_2^\pi)^{+1.4}_{-1.1}(B_{1,3}^{S})^{+0.4}_{-0.7}(m_0)$
   \\
$\rm{B_c \to a_0^0 \pi^+}$ & $3.5^{+1.6}_{-1.0}(m_c)^{+0.4}_{-0.4}(\bar f_S)^{+1.0}_{-0.6}(a_2^\pi)^{+1.1}_{-0.9}(B_{1,3}^{S})^{+0.7}_{-0.4}(m_0)$
   \\
     \hline
 $\rm{B_c \to a_0^+ \eta}\times 10$  &$3.6^{+3.4}_{-0.9}(m_c)^{+0.4}_{-0.3}(\bar f_S)^{+1.7}_{-0.4}(a_2^\eta)^{+1.6}_{-0.6}(B_{1,3}^{S})^{+0.0}_{-0.0}(m_0)$
 \\
 $\rm{B_c \to a_0^+ \eta'}\times 10$ & $2.4^{+2.2}_{-0.6}(m_c)^{+0.3}_{-0.2}(\bar f_S)^{+1.1}_{-0.3}(a_2^{\eta'})^{+1.1}_{-0.4}(B_{1,3}^{S})^{+0.0}_{-0.0}(m_0)$
\\
  \hline
$\rm{B_c \to  \overline{\kappa}^0 K^+ }$   & $3.4^{+2.1}_{-1.1}(m_c)^{+0.5}_{-0.4}(\bar f_S)^{+1.8}_{-1.4}(a_{1,2}^K)^{+1.4}_{-0.9}(B_{1,3}^{S})^{+0.2}_{-0.0}(m_0)$
\\
$\rm{B_c \to \overline{K}^0 \kappa^+  }$ & $2.1^{+0.1}_{-0.0}(m_c)^{+0.3}_{-0.2}(\bar f_S)^{+1.5}_{-0.1}(a_{1,2}^K)^{+0.7}_{-0.4}(B_{1,3}^{S})^{+0.3}_{-0.1}(m_0)$
\\
  \hline
$\rm{B_c \to \pi^+ \sigma }\times 10$ & $3.2^{+1.9}_{-0.0}(m_c)^{+0.3}_{-0.3}(\bar f_S)^{+1.1}_{-1.3}(a_2^\pi)^{+0.9}_{-0.7}(B_{1,3}^{S})^{+0.1}_{-0.3}(m_0) (\rm{f_0^q})$
   \\
$\rm{B_c \to \pi^+ f_0 }\times 10$ & $1.8^{+1.1}_{-0.0}(m_c)^{+0.2}_{-0.2}(\bar f_S)^{+1.6}_{-0.6}(a_2^\pi)^{+0.7}_{-0.2}(B_{1,3}^{S})^{+0.7}_{-0.0}(m_0) (\rm{f_0^q})$    \\
\hline \hline
\end{tabular}
\end{center}
\end{table}

In this subsection, we will make some discussions on the $B_c \to a (P, V)$
decays  involving 14 $\Delta S =0$ and 8 $\Delta S =1$ processes, respectively.

From the numerical results for considered modes as given in the Tables~\ref{tab:bcsp-1s0}, \ref{tab:bcsp-2s0},
\ref{tab:bcsv-1s0}, and \ref{tab:bcsv-2s0}, one can find that the {\it CP-}averaged BRs for all the
$\Delta S =0 $ $B_c \to a (P, V)$ processes are
in the range of $10^{-6} \sim 10^{-5}$ within the theoretical errors except for $B_c \to a_0^+ \eta^{(')}$ decays,
which are expected to be tested by the
ongoing LHCb measurements and the forthcoming Super-B experiments. Since we make the perturbative calculations
based on the assumption of two-quark structure for the scalars, once these theoretical predictions could
be verified by the related experiments, then these results will
help us to explore the underlying structure of the scalar $a$ meson.

For $B_c \to a_0 (\pi, \rho)$ decays,
their BRs can be read from the
Tables~\ref{tab:bcsp-1s0} and \ref{tab:bcsv-1s0} (in unit of $10^{-6}$),
\beq
Br(B_c \to a_0^+  \pi^0) &=& 6.5^{+3.6}_{-2.5}    \;, \quad \;
Br(B_c \to a_0^0  \pi^+) = 3.5^{+2.3}_{-1.6}      \;, \label{eq:bra0pi}\\
Br(B_c \to a_0^+  \rho^0) &=& 12.7^{+6.1}_{-5.6}     \;, \quad
Br(B_c \to a_0^0  \rho^+) = 10.6^{+5.5}_{-3.5}    \;, \label{eq:bra0rho}
\eeq
where the various errors as specified have been added in quadrature. One could find the rather different
decay patterns from these
theoretical predictions, i.e., Eqs.~(\ref{eq:bra0pi}, \ref{eq:bra0rho}) that $Br(B_c \to a_0^+  \pi^0) >
 Br(B_c \to a_0^0  \pi^+)$ while
 $Br(B_c \to a_0^+  \rho^0) \sim
Br(B_c \to a_0^0  \rho^+)$ within the theoretical uncertainties. Because $f_\rho(f_\rho^T) \sim 1.6(1.3) \times f_\pi$, it is evident
that $Br(B_c \to a_0 \rho) > Br(B_c \to a_0 \pi)$.
Based on these pQCD predictions of BRs for $B_c \to a_0 (\pi, \rho)$ decays, which are within the reach of
LHCb experiments~\cite{ekou09:ncbc}, it is expected that if the observation or the
experimental upper limit on the decay modes $B_c \to a_0 \pi (a_0 \rho)$
 are much smaller than the expectation,
 this might rule out the $q\bar q$ structure for
the $a_0$.

On the other hand,
the isovector scalar meson $a_0(1450)$ has been
confirmed to be a conventional $q\bar{q}$ meson in lattice
calculations~\cite{Mathur06:lattice,Kim97-Burch06:lattice,Bardeen02:lattice,Kunihiro04:lattice,Prelovsek04:lattice} recently. Hence, the
calculations for the $a_0(1450)$ channels should be more
trustworthy. Our results shown in Tables~\ref{tab:bcsp-2s0} and \ref{tab:bcsv-2s0} indicate that $B_c \to a_0(1450) \pi$ and
$B_c \to a_0(1450) \rho$ have large branching ratios, of order
$(5 \sim 20)\times 10^{-6}$ and $(15 \sim 47)\times 10^{-6}$,
respectively.
A measurement of them at the predicted level will reinforce the
$q\bar q$ nature for the $a_0(1450)$.


\begin{table}[t]
\caption{ Same as Table~\ref{tab:bcsp-1s0} but for the $\Delta S =1$ processes
of charmless hadronic $B_c \to
(a_0, \kappa, \sigma, f_0)(\pi, K, \eta, \eta')$ decays in S1.}
\label{tab:bcsp-1s1}
\begin{center}\vspace{-0.6cm}
\begin{tabular}[t]{l|l}
 \hline \hline
 $\Delta S =1$ &                      \\
 Decay modes   &    BRs $(10^{-7})$    \\
\hline
$\rm{B_c \to a_0^+ K^0}$         &$4.0^{+0.4}_{-0.9}(m_c)^{+0.4}_{-0.5}(\bar f_S)^{+1.7}_{-1.7}(a_{1,2}^K)^{+0.7}_{-1.0}(B_{1,3}^{S})^{+0.0}_{-0.2}(m_0)$
   \\
$\rm{B_c \to a_0^0 K^+}$         &$2.0^{+0.2}_{-0.5}(m_c)^{+0.2}_{-0.3}(\bar f_S)^{+0.9}_{-0.9}(a_{1,2}^K)^{+0.4}_{-0.5}(B_{1,3}^{S})^{+0.0}_{-0.1}(m_0)$
   \\
     \hline
 $\rm{B_c \to \kappa^+ \eta}$          & $4.5^{+1.3}_{-0.9}(m_c)^{+0.5}_{-0.5}(\bar f_S)^{+0.9}_{-0.6}(a_2^\eta)^{+0.8}_{-0.9}(B_{1,3}^{S})^{+0.0}_{-0.0}(m_0)$
 \\
$\rm{B_c \to \kappa^+ \eta'}\times 10$ & $8.8^{+2.4}_{-0.1}(m_c)^{+1.3}_{-0.8}(\bar f_S)^{+0.6}_{-0.7}(a_2^{\eta'})^{+3.7}_{-2.5}(B_{1,3}^{S})^{+0.0}_{-0.0}(m_0)$
\\
  \hline
$\rm{B_c \to \kappa^0 \pi^+}$ & $2.1^{+1.1}_{-0.6}(m_c)^{+0.2}_{-0.2}(\bar f_S)^{+0.6}_{-0.3}(a_2^\pi)^{+0.6}_{-0.4}(B_{1,3}^{S})^{+0.1}_{-0.0}(m_0)$
\\
$\rm{B_c \to  \kappa^+ \pi^0}$& $1.1^{+0.6}_{-0.3}(m_c)^{+0.1}_{-0.1}(\bar f_S)^{+0.3}_{-0.2}(a_2^\pi)^{+0.4}_{-0.2}(B_{1,3}^{S})^{+0.1}_{-0.0}(m_0)$
\\
  \hline
$\rm{B_c \to K^+ \sigma }$&$1.6^{+0.2}_{-0.3}(m_c)^{+0.1}_{-0.2}(\bar f_S)^{+0.6}_{-0.7}(a_{1,2}^K)^{+0.4}_{-0.4}(B_{1,3}^{S})^{+0.0}_{-0.2}(m_0) (\rm{f_0^q})$
   \\
                           &$0.9^{+0.5}_{-0.5}(m_c)^{+0.1}_{-0.1}(\bar f_S)^{+0.3}_{-0.5}(a_{1,2}^K)^{+0.2}_{-0.4}(B_{1,3}^{S})^{+0.0}_{-0.1}(m_0) (\rm{f_0^s})$   \\
$\rm{B_c \to K^+ f_0}$&$1.8^{+0.4}_{-0.3}(m_c)^{+0.2}_{-0.2}(\bar f_S)^{+0.6}_{-0.6}(a_{1,2}^K)^{+0.3}_{-0.4}(B_{1,3}^{S})^{+0.0}_{-0.0}(m_0) (\rm{f_0^q})$
   \\
                           &$0.3^{+0.5}_{-0.2}(m_c)^{+0.0}_{-0.1}(\bar f_S)^{+0.4}_{-0.2}(a_{1,2}^K)^{+0.1}_{-0.1}(B_{1,3}^{S})^{+0.0}_{-0.0}(m_0) (\rm{f_0^s})$   \\
\hline \hline
\end{tabular}
\end{center}
\end{table}

For those $B_c \to a (P, V)$ decay modes with $a_0(1450)$ as one of the final states, the pQCD predictions
 in Tables~\ref{tab:bcsp-2s0} and \ref{tab:bcsv-2s0} show that for the $\Delta S =0$ processes $B_c \to a_0(1450) (\pi, \eta^{(')}, \rho, \omega)$
 the BRs in S1 are much larger than that in S2, however,
 for the $\Delta S =1$ processes $B_c \to a_0(1450) K^{(*)}$, the BRs in S1 are much smaller than that in S2, which will be confronted with
 the ongoing and forthcoming related experiments. It is hoped that the precision measurements could help us to determine which scenario
 is favored by the experiments, then the inner quark structure definitely.

For $B_c \to a (\eta, \eta')$ decays, the numerical results grouped in the Tables~\ref{tab:bcsp-1s0} and \ref{tab:bcsp-2s0} indicate
the small differences between $B_c \to a \eta$ and $B_c \to a \eta'$ modes, which is mainly because
the relevant final state mesons, $\eta^{(')}$, contain the same component $\bar u u + \bar d d$, just
with the different coefficients, i.e., $\cos\phi$ and $\sin\phi$.
This pattern is very similar to that of $B_c \to \rho \eta^{(')}$ decays~\cite{Xiao1:bcdecay}.

For the $\Delta S =1$ $B_c \to a K^{(*)}$ processes, the pQCD predicted BRs are in the order of $10^{-7}$, which is below the reach of
the LHCb experiments($\sim 10^{-6}$). From these numerical results as displayed in Tables~\ref{tab:bcsp-1s1}, \ref{tab:bcsp-2s1},
\ref{tab:bcsv-1s1}, and \ref{tab:bcsv-2s1}, one can find that $Br(B_c \to a^+ K^{(*)0}) \approx 2 \times Br(B_c \to a^0 K^{(*)+})$
 although for $a^0$ meson the vector decay constant $f_{a^0}=0$, which exhibits clearly that the contribution is dominated by the odd Gegenbauer moments
  in the leading twist distribution amplitude of the scalar $a$ meson.
 This pattern is well consistent with that 
 stressed by the authors in Ref.~\cite{Cheng06:B2SP}.


\begin{table}[htb]
\caption{ Same as Table~\ref{tab:bcsp-1s0} but for the $\Delta S = 0$ processes of
charmless hadronic $B_c \to (a_0(1450), {K_0^*}(1430), f_0(1370), f_0(1500))(\pi, K, \eta, \eta')$
decays in S1 and S2, respectively.}
\label{tab:bcsp-2s0}
\begin{center}\vspace{-0.6cm}
\begin{tabular}[t]{l|l}
 \hline \hline
 $\Delta S =0$     &                                      \\
 Decay modes       &    BRs $(10^{-6})$      \\
\hline
$\rm{B_c \to a_0(1450)^+ \pi^0}$              &$21.0^{+6.9}_{-5.7}(m_c)^{+4.7}_{-4.3}(\bar f_S)^{+4.9}_{-5.6}(a_2^\pi)^{+4.0}_{-5.7}(B_{1,3}^{S})^{+0.0}_{-0.2}(m_0)$ (S1) \\
                                              &$6.3^{+4.4}_{-2.6}(m_c)^{+1.4}_{-1.3}(\bar f_S)^{+0.8}_{-0.3}(a_2^\pi)^{+2.8}_{-1.8}(B_{1,3}^{S})^{+0.5}_{-0.2}(m_0)$ (S2)\\
$\rm{B_c \to a_0(1450)^0 \pi^+}$             &$11.9^{+3.8}_{-3.1}(m_c)^{+2.7}_{-2.4}(\bar f_S)^{+1.3}_{-2.4}(a_2^\pi)^{+2.8}_{-2.1}(B_{1,3}^{S})^{+0.5}_{-0.9}(m_0)$ (S1) \\
                                             &$4.9^{+3.6}_{-2.3}(m_c)^{+1.1}_{-1.0}(\bar f_S)^{+0.4}_{-0.4}(a_2^\pi)^{+2.6}_{-1.2}(B_{1,3}^{S})^{+0.0}_{-0.4}(m_0)$ (S2)
\\
  \hline
$\rm{B_c \to a_0(1450)^+ \eta}$              &$2.7^{+0.4}_{-0.1}(m_c)^{+0.6}_{-0.4}(\bar f_S)^{+0.5}_{-0.0}(a_2^\eta)^{+1.1}_{-0.7}(B_{1,3}^{S})^{+0.0}_{-0.0}(m_0)$ (S1)  \\
                                             &$1.0^{+0.2}_{-0.3}(m_c)^{+0.2}_{-0.2}(\bar f_S)^{+0.2}_{-0.1}(a_2^\eta)^{+0.7}_{-0.5}(B_{1,3}^{S})^{+0.0}_{-0.0}(m_0)$ (S2)\\
$\rm{B_c \to a_0(1450)^+ \eta'}$             &$1.1^{+0.2}_{-0.1}(m_c)^{+0.4}_{-0.3}(\bar f_S)^{+0.3}_{-0.0}(a_2^{\eta'})^{+0.7}_{-0.5}(B_{1,3}^{S})^{+0.0}_{-0.0}(m_0)$ (S1)  \\
                                             &$0.6^{+0.2}_{-0.2}(m_c)^{+0.2}_{-0.1}(\bar f_S)^{+0.1}_{-0.1}(a_2^{\eta'})^{+0.5}_{-0.3}(B_{1,3}^{S})^{+0.0}_{-0.0}(m_0)$ (S2)
 \\
   \hline
$\rm{B_c \to \overline{K}_0^*(1430)^0 K^+}$         &$19.2^{+4.7}_{-5.4}(m_c)^{+4.1}_{-3.9}(\bar f_S)^{+2.5}_{-3.8}(a_{1,2}^K)^{+2.2}_{-1.9}(B_{1,3}^{S})^{+0.3}_{-0.5}(m_0)$ (S1) \\
                                                    &$0.7^{+0.7}_{-0.2}(m_c)^{+0.2}_{-0.2}(\bar f_S)^{+0.8}_{-0.5}(a_{1,2}^K)^{+1.1}_{-0.9}(B_{1,3}^{S})^{+0.0}_{-0.1}(m_0)$ (S2) \\
$\rm{B_c \to \overline{K}^0 {K_0^*}(1430)^+}$       &$4.3^{+0.8}_{-0.9}(m_c)^{+0.9}_{-0.9}(\bar f_S)^{+1.4}_{-2.2}(a_{1,2}^K)^{+1.0}_{-0.8}(B_{1,3}^{S})^{+0.2}_{-0.0}(m_0)$ (S1)\\
                                                    &$9.2^{+1.4}_{-2.0}(m_c)^{+2.0}_{-2.0}(\bar f_S)^{+0.7}_{-0.9}(a_{1,2}^K)^{+3.8}_{-4.3}(B_{1,3}^{S})^{+0.0}_{-0.4}(m_0)$ (S2)\\
  \hline
$\rm{B_c \to f_0(1370) \pi^+}$                &$3.6^{+1.2}_{-0.7}(m_c)^{+0.9}_{-0.8}(\bar f_S)^{+1.9}_{-1.3}(a_2^\pi)^{+0.9}_{-1.2}(B_{1,3}^{S})^{+0.3}_{-0.0}(m_0) (\rm{f_0^q, S1})$\\
                                              &$1.0^{+0.2}_{-0.3}(m_c)^{+0.2}_{-0.2}(\bar f_S)^{+0.4}_{-0.4}(a_2^\pi)^{+1.0}_{-0.9}(B_{1,3}^{S})^{+0.1}_{-0.1}(m_0) (\rm{f_0^q, S2})$\\
$\rm{B_c \to f_0(1500) \pi^+}$               &$3.7^{+1.1}_{-1.0}(m_c)^{+0.9}_{-0.8}(\bar f_S)^{+1.8}_{-1.5}(a_2^\pi)^{+0.6}_{-1.3}(B_{1,3}^{S})^{+0.2}_{-0.3}(m_0) (\rm{f_0^q, S1})$ \\
                                             &$0.9^{+0.2}_{-0.3}(m_c)^{+0.2}_{-0.2}(\bar f_S)^{+0.5}_{-0.2}(a_2^\pi)^{+1.1}_{-0.6}(B_{1,3}^{S})^{+0.1}_{-0.0}(m_0) (\rm{f_0^q, S2})$
 \\
\hline \hline
\end{tabular}
\end{center}
\end{table}

\begin{table}[htb]
\caption{ Same as Table~\ref{tab:bcsp-1s0} but for the $\Delta S = 1$ processes of
charmless hadronic $B_c \to (a_0(1450), {K_0^*}(1430), f_0(1370), f_0(1500))(\pi, K, \eta, \eta')$
decays in S1 and S2, respectively.}
\label{tab:bcsp-2s1}
\begin{center}\vspace{-0.6cm}
\begin{tabular}[t]{l|l}
 \hline \hline
$\Delta S =1$ &                      \\
Decay modes   &    BRs $(10^{-7})$    \\
\hline
$\rm{B_c \to a_0(1450)^+ K^0}$              &$2.3^{+1.0}_{-0.7}(m_c)^{+0.5}_{-0.5}(\bar f_S)^{+2.3}_{-1.2}(a_{1,2}^K)^{+1.2}_{-0.8}(B_{1,3}^{S})^{+0.1}_{-0.1}(m_0)$ (S1)\\
                                            &$6.7^{+2.6}_{-1.8}(m_c)^{+1.6}_{-1.4}(\bar f_S)^{+2.3}_{-0.8}(a_{1,2}^K)^{+2.5}_{-1.9}(B_{1,3}^{S})^{+0.1}_{-0.2}(m_0)$ (S2) \\
$\rm{B_c \to a_0(1450)^0 K^+}$               &$1.2^{+0.5}_{-0.4}(m_c)^{+0.3}_{-0.3}(\bar f_S)^{+1.2}_{-0.6}(a_{1,2}^K)^{+0.6}_{-0.4}(B_{1,3}^{S})^{+0.1}_{-0.1}(m_0)$ (S1) \\ 
                                             &$3.4^{+1.3}_{-0.9}(m_c)^{+0.8}_{-0.7}(\bar f_S)^{+1.2}_{-0.4}(a_{1,2}^K)^{+1.3}_{-1.0}(B_{1,3}^{S})^{+0.1}_{-0.1}(m_0)$ (S2)
\\
  \hline
$\rm{B_c \to {K_0^*}(1430)^+ \eta}$          &$5.4^{+2.8}_{-1.9}(m_c)^{+1.1}_{-1.1}(\bar f_S)^{+1.5}_{-1.6}(a_2^\eta)^{+0.7}_{-0.8}(B_{1,3}^{S})^{+0.0}_{-0.0}(m_0)$ (S1) \\
                                             &$6.2^{+2.6}_{-1.9}(m_c)^{+1.5}_{-1.3}(\bar f_S)^{+0.3}_{-0.0}(a_2^\eta)^{+1.4}_{-2.5}(B_{1,3}^{S})^{+0.0}_{-0.0}(m_0)$ (S2) \\
$\rm{B_c \to {K_0^*}(1430)^+ \eta'}$         &$3.3^{+0.0}_{-0.6}(m_c)^{+0.5}_{-0.7}(\bar f_S)^{+0.2}_{-0.3}(a_2^{\eta'})^{+0.0}_{-0.5}(B_{1,3}^{S})^{+0.0}_{-0.0}(m_0)$ (S1) \\
                                             &$5.1^{+0.2}_{-0.3}(m_c)^{+1.2}_{-1.1}(\bar f_S)^{+0.7}_{-0.3}(a_2^{\eta'})^{+2.7}_{-1.7}(B_{1,3}^{S})^{+0.0}_{-0.0}(m_0)$ (S2)
 \\
   \hline
$\rm{B_c \to {K_0^*}(1430)^0 \pi^+}$          &$6.5^{+3.5}_{-2.3}(m_c)^{+1.2}_{-1.3}(\bar f_S)^{+0.4}_{-0.8}(a_2^\pi)^{+0.9}_{-0.8}(B_{1,3}^{S})^{+0.1}_{-0.2}(m_0)$ (S1) \\
                                              &$0.8^{+0.9}_{-0.5}(m_c)^{+0.2}_{-0.2}(\bar f_S)^{+0.0}_{-0.1}(a_2^\pi)^{+0.5}_{-0.4}(B_{1,3}^{S})^{+0.0}_{-0.1}(m_0)$ (S2) \\
$\rm{B_c \to {K_0^*}(1430)^+ \pi^0}$         &$3.2^{+1.8}_{-1.2}(m_c)^{+0.6}_{-0.6}(\bar f_S)^{+0.2}_{-0.4}(a_2^\pi)^{+0.5}_{-0.4}(B_{1,3}^{S})^{+0.1}_{-0.1}(m_0)$ (S1) \\
                                             &$0.4^{+0.5}_{-0.3}(m_c)^{+0.1}_{-0.1}(\bar f_S)^{+0.0}_{-0.1}(a_2^\pi)^{+0.3}_{-0.2}(B_{1,3}^{S})^{+0.0}_{-0.1}(m_0)$ (S2) \\
  \hline
$\rm{B_c \to f_0(1370) K^+}$              &$0.9^{+0.4}_{-0.3}(m_c)^{+0.3}_{-0.2}(\bar f_S)^{+0.9}_{-0.4}(a_{1,2}^K)^{+0.5}_{-0.3}(B_{1,3}^{S})^{+0.0}_{-0.0}(m_0) (\rm{f_0^q, S1})$ \\
                                          &$2.7^{+0.9}_{-0.8}(m_c)^{+0.6}_{-0.5}(\bar f_S)^{+0.6}_{-0.3}(a_{1,2}^K)^{+1.2}_{-1.1}(B_{1,3}^{S})^{+0.1}_{-0.1}(m_0) (\rm{f_0^q, S2})$ \\
                                              &$7.7^{+1.8}_{-1.7}(m_c)^{+1.9}_{-1.7}(\bar f_S)^{+3.5}_{-2.2}(a_{1,2}^K)^{+1.6}_{-2.0}(B_{1,3}^{S})^{+0.1}_{-0.0}(m_0) (\rm{f_0^s, S1})$\\
                                              &$0.7^{+0.2}_{-0.0}(m_c)^{+0.1}_{-0.2}(\bar f_S)^{+0.8}_{-0.4}(a_{1,2}^K)^{+0.6}_{-0.5}(B_{1,3}^{S})^{+0.0}_{-0.0}(m_0) (\rm{f_0^s, S2})$\\
$\rm{B_c \to f_0(1500) K^+}$           &$0.9^{+0.4}_{-0.3}(m_c)^{+0.2}_{-0.2}(\bar f_S)^{+0.8}_{-0.4}(a_{1,2}^K)^{+0.5}_{-0.3}(B_{1,3}^{S})^{+0.0}_{-0.0}(m_0) (\rm{f_0^q, S1})$ \\
                                       &$2.8^{+1.0}_{-0.8}(m_c)^{+0.5}_{-0.6}(\bar f_S)^{+0.4}_{-0.4}(a_{1,2}^K)^{+1.1}_{-1.1}(B_{1,3}^{S})^{+0.1}_{-0.1}(m_0) (\rm{f_0^q, S2})$  \\
                                             &$7.9^{+1.5}_{-1.3}(m_c)^{+2.0}_{-1.7}(\bar f_S)^{+3.5}_{-2.3}(a_{1,2}^K)^{+1.6}_{-2.0}(B_{1,3}^{S})^{+0.4}_{-0.0}(m_0) (\rm{f_0^s, S1})$\\
                                             &$0.8^{+0.2}_{-0.0}(m_c)^{+0.2}_{-0.1}(\bar f_S)^{+0.8}_{-0.4}(a_{1,2}^K)^{+0.5}_{-0.5}(B_{1,3}^{S})^{+0.0}_{-0.0}(m_0) (\rm{f_0^s, S2})$ \\
\hline \hline
\end{tabular}
\end{center}
\end{table}

\subsection{ $B_c \to \kappa (P, V)$ and $B_c \to K_0^*(1430) (P, V)$ decays}

In this type of the considered decays, there are 8 $B_c \to K_0^* K^{(*)}(\Delta S=0)$ modes and
16 $B_c \to K_0^* (\pi, \eta^{(')},\rho, \omega, \phi)(\Delta S=1)$ channels.

In the $\Delta S =0$ processes, we have 4 $B_c \to \kappa^+ \overline{K}^{(*)0},~ \overline{\kappa}^0 K^{(*)+}$ channels in S1
and 4 $B_c \to K_0^*(1430)^+ \overline{K}^{(*)0},~ \overline{K}_0^*(1430)^0 K^{(*)+}$ decays in both S1 and S2, respectively.
From the pQCD predictions for these considered modes as given in the Tables~\ref{tab:bcsp-1s0}, \ref{tab:bcsp-2s0},
\ref{tab:bcsv-1s0}, and \ref{tab:bcsv-2s0}, one can observe that all the BRs are in the range of $10^{-6} \sim 10^{-5}$ within the theoretical errors,
which could be measured by the near future LHCb 
and Super-B experiments operated at CERN and KEK, respectively.

Here, it is very interesting to note that for $B_c \to K_0^* K^{(*)}$ channels
$Br(B_c \to \overline{K}_0^{*0} K^{+}) > Br(B_c \to {K_0^*}^+ \overline{K}^{0})$ and
$Br(B_c \to \overline{K}_0^{*0} K^{*+}) > Br(B_c \to {K_0^*}^+ \overline{K}^{*0})$ in S1, respectively, while the situation
is quite the contrary for $B_c \to K_0^*(1430) K^{(*)}$ decays in S2. One can also find that for the $B_c \to \overline{K}_0^*(1430)^0 K^{(*)+}$
decays in both scenarios $Br(B_c \to \overline{K}_0^*(1430)^0 K^{(*)+})_{{\rm S1}} >> Br(B_c \to \overline{K}_0^*(1430)^0 K^{(*)+})_{{\rm S2}}$, while
for the $B_c \to K_0^*(1430)^+ \overline{K}^{(*)0}$ modes,
$Br(B_c \to K_0^*(1430)^+ \overline{K}^{(*)0})_{{\rm S1}} < Br(B_c \to K_0^*(1430)^+ \overline{K}^{(*)0})_{{\rm S2}}$.
It should be stressed that once these predicted BRs and the relevant relations could be tested by the experiments in the near future,
this could provide the great opportunities for us to
explore the physical properties of the scalars $K_0^*$ and the corresponding annihilation decay mechanism.

For the $\Delta S =1$ channels $B_c \to K_0^* (\pi, \eta^{(')})$ and $B_c \to K_0^* (\rho, \omega, \phi)$,
all the theoretical BRs in the pQCD approach are in the
range of $10^{-8} \sim 10^{-7}$ within the theoretical errors
except for $Br(B_c \to K_0^*(1430) (\rho, \omega))_{\rm{S1}} \sim 10^{-6}$
though they
are CKM
suppressed($V_{us}= 0.22$),
which will be confronted by the ongoing and forthcoming relevant experimental measurements. Due to
the contributions from the same component, i.e., $u \bar u$, and few differences of the decay constants and masses
between $\rho^0$ and $\omega$,
which result in the similar BRs for $B_c \to K^*_0 \rho^0$ and $B_c \to K_0^* \omega$ in the considered scenarios. Moreover, we find that
the simple relations $Br(B_c \to {K^*_0}^0 (\pi^+, \rho^+)) \approx 2 \times Br(B_c \to {K^*_0}^+ (\pi^0, \rho^0))$
exists in our pQCD perturbative calculations exactly and $Br(B_c \to K^*_0(1430) (\pi, \rho, \omega))_{{\rm S1}} >
Br(B_c \to K^*_0(1430) (\pi, \rho, \omega))_{{\rm S2}}$. However, $Br(B_c \to K^*_0(1430)^+ \phi)_{{\rm S1}} <
Br(B_c \to K^*_0(1430)^+ \phi)_{{\rm S2}}$, whose pattern agrees well with that obtained by Kim, Li, and Wang in Ref.~\cite{Wang:b2scalar}.

For $B_c \to {K_0^*}^+ (\eta, \eta')$ decay modes, based on the pQCD numerical results, we have the following remarks:
In this sector, both of the components $\eta_q$ and $\eta_s$ in $\eta$ and $\eta'$ contribute to these channels but with
different coefficients even
opposite sign. For $B_c \to \kappa^+ \eta^{(')}$ decays, the two parts of contributions make a
constructive interference to the branching ratio of $B_c \to \kappa^+ \eta$,
while a destructive interference to that of $B_c \to  \kappa^+ \eta'$, which eventually
results in $Br(B_c \to \kappa^+ \eta) \approx  5 \times Br(B_c \to  \kappa^+ \eta')$. This pattern is very like
that of $B \to K^{*} \eta$ and $K^{*} \eta'$ decay channels~\cite{Amsler08:pdg,Barberio08:hfag}. For $B_c \to K_0^*(1430)^+ \eta^{(')}$ modes,
unlike the $B_c \to \kappa^+ \eta^{(')}$, both of them are determined mainly by the factorizable contributions of $\eta_s$ term,
which leads to $Br(B_c \to K_0^*(1430)^+ \eta) \sim Br(B_c \to K_0^*(1430)^+ \eta^{'})$ within the theoretical errors in both
scenarios. Meanwhile, it is interesting to note that 
$Br(B_c \to K_0^*(1430)^+ \eta^{(')})_{{\rm S1}}
 < Br(B_c \to K_0^*(1430)^+ \eta^{(')})_{{\rm S2}}$ while $Br(B_c \to  K_0^*(1430)^+ \eta) > Br(B_c \to K_0^*(1430)^+ \eta^{'})$ in both scenarios,
 where only the central values are quoted for comparison.
Because of the small BRs($< 10^{-6}$) for $B_c \to {K_0^*}^+ \eta^{(')}$ decays, all the above theoretical pQCD predictions
of the BRs and the physical relations are expected to be examined in the forthcoming 
Super-B experiments.


\begin{table}[t]
\caption{ Same as Table~\ref{tab:bcsp-1s0} but for the $\Delta S = 0$ processes of
charmless hadronic $B_c \to (a_0, \kappa, \sigma, f_0)(\rho, K^*, \omega, \phi)$ decays in S1.}
\label{tab:bcsv-1s0}
\begin{center}\vspace{-0.6cm}
\begin{tabular}[t]{l|l}
 \hline \hline
 $\Delta S =0$     &                                 \\
 Decay modes       &    BRs $(10^{-6})$              \\
\hline
$\rm{B_c \to a_0^+ \rho^0}$             & $12.7^{+4.4}_{-3.8}(m_c)^{+1.5}_{-1.3}(\bar f_S)^{+2.9}_{-2.9}(a_2^\rho)^{+2.7}_{-2.7}(B_{1,3}^{S})$
   \\
$\rm{B_c \to a_0^0 \rho^+}$             & $10.6^{+4.4}_{-2.6}(m_c)^{+1.2}_{-1.1}(\bar f_S)^{+1.5}_{-1.0}(a_2^\rho)^{+2.7}_{-1.9}(B_{1,3}^{S})$
   \\
 $\rm{B_c \to a_0^+ \omega} \times 10$  & $9.8^{+9.2}_{-3.0}(m_c)^{+0.8}_{-1.3}(\bar f_S)^{+3.3}_{-1.9}(a_2^\omega)^{+3.5}_{-1.7}(B_{1,3}^{S})$
\\
  \hline
$\rm{B_c \to  \overline{\kappa}^0 K^{*+} }$   & $8.8^{+4.5}_{-2.5}(m_c)^{+1.1}_{-0.9}(\bar f_S)^{+2.2}_{-1.0}(a_{1,2}^{K^*})^{+3.0}_{-1.7}(B_{1,3}^{S})$
\\
$\rm{B_c \to \overline{K}^{*0} \kappa^+ }$  & $4.9^{+0.8}_{-0.7}(m_c)^{+0.6}_{-0.5}(\bar f_S)^{+0.6}_{-1.0}(a_{1,2}^{K^*})^{+1.7}_{-1.6}(B_{1,3}^{S})$
\\
  \hline
$\rm{B_c \to \rho^+  \sigma }\times 10$      & $1.6^{+2.1}_{-0.0}(m_c)^{+0.2}_{-0.2}(\bar f_S)^{+1.5}_{-1.1}(a_2^\rho)^{+0.6}_{-0.9}(B_{1,3}^{S}) (\rm{f_0^q})$
   \\
$\rm{B_c \to \rho^+ f_0}\times 10$      & $0.8^{+1.4}_{-0.0}(m_c)^{+0.1}_{-0.1}(\bar f_S)^{+0.6}_{-0.6}(a_2^\rho)^{+0.7}_{-0.5}(B_{1,3}^{S}) (\rm{f_0^q})$
   \\
\hline \hline
\end{tabular}
\end{center}
\end{table}

\subsection{ $B_c \to f (P, V)$ and $B_c \to f' (P, V)$ decays}

As mentioned in the above sections, it is well known that the identification of the structure
of these neutral scalar mesons $f$ and $f'$ is very difficult, which is a longstanding puzzle
not yet resolved either by experimentalists or by theorists. Although various scenarios on their component
have been proposed, by considering the feasibility of factorization approach, we here
assume
these considered scalars to be only $q\bar q$ bound states.

For the considered 16 $B_c \to (f, f') (P, V)$ decays, the numerical pQCD predictions have been displayed
in the Tables~\ref{tab:bcsp-1s0}-\ref{tab:bcsv-2s1}. For the $f$ and $f'$, the quarkonia component has been
proposed, which can be seen in Eqs.~(\ref{eq:f0mixing}) and~(\ref{eq:f0mixing1}).
For the $\Delta S =0$ processes $B_c \to (f, f')(\pi^+, \rho^+)$,
we use the pure $q \bar q$ states $f_0^q$ in $f$ and $f'$ to calculate the BRs in the pQCD
 approach and obtain the numerical results, in which one can find that
the BRs of $B_c \to (\pi^+, \rho^+) (f_0(1370), f_0(1500))(q\bar q)$ are in the order of
$10^{-6}$ within the theoretical errors in both scenarios and within the reach of the LHCb experiments~\cite{ekou09:ncbc}, while
the BRs of $B_c \to (\pi^+, \rho^+) (\sigma, f_0)(q\bar q)$ are highly below the experimental reach of LHCb at CERN.
 Here, we have assumed that $\sigma$ and $f_0(1370)$ have the similar decay constant and light-cone distribution amplitudes
as $f_0$ and $f_0(1500)$, respectively.


\begin{table}[t]
\caption{ Same as Table~\ref{tab:bcsp-1s0} but for the $\Delta S = 1$ processes of
charmless hadronic $B_c \to (a_0, \kappa, \sigma, f_0)(\rho, K^*, \omega, \phi)$ decays in S1.}
\label{tab:bcsv-1s1}
\begin{center}\vspace{-0.6cm}
\begin{tabular}[t]{l|l}
 \hline \hline
 $\Delta S =1$ &                      \\
 Decay modes   &    BRs $(10^{-7})$    \\
\hline
$\rm{B_c \to a_0^+ K^{*0}}$       &$3.5^{+0.8}_{-0.4}(m_c)^{+0.4}_{-0.3}(\bar f_S)^{+0.8}_{-0.5}(a_{1,2}^{K^*})^{+1.2}_{-0.8}(B_{1,3}^{S})$
   \\
$\rm{B_c \to a_0^0 K^{*+}}$       &$1.7^{+0.4}_{-0.2}(m_c)^{+0.2}_{-0.2}(\bar f_S)^{+0.4}_{-0.3}(a_{1,2}^{K^*})^{+0.6}_{-0.4}(B_{1,3}^{S})$
   \\
     \hline
 $\rm{B_c \to \kappa^+ \omega}$  &$1.9^{+1.0}_{-0.6}(m_c)^{+0.2}_{-0.2}(\bar f_S)^{+0.2}_{-0.2}(a_2^\omega)^{+0.6}_{-0.5}(B_{1,3}^{S})$
 \\
$\rm{B_c \to \kappa^+ \phi}$     &$2.9^{+0.7}_{-0.6}(m_c)^{+0.3}_{-0.3}(\bar f_S)^{+0.2}_{-0.6}(a_2^\phi)^{+0.9}_{-1.0}(B_{1,3}^{S})$
\\
  \hline
$\rm{B_c \to \kappa^0 \rho^+}$   &$4.5^{+2.4}_{-1.4}(m_c)^{+0.5}_{-0.5}(\bar f_S)^{+0.7}_{-0.3}(a_2^\rho)^{+1.4}_{-0.9}(B_{1,3}^{S})$
\\
$\rm{B_c \to  \kappa^+ \rho^0}$  &$2.3^{+1.2}_{-0.7}(m_c)^{+0.3}_{-0.3}(\bar f_S)^{+0.4}_{-0.2}(a_2^\rho)^{+0.8}_{-0.4}(B_{1,3}^{S})$
\\
  \hline
$\rm{B_c \to K^{*+} \sigma }$         &$1.6^{+0.3}_{-0.1}(m_c)^{+0.2}_{-0.2}(\bar f_S)^{+0.3}_{-0.2}(a_{1,2}^{K^*})^{+0.4}_{-0.4}(B_{1,3}^{S}) (\rm{f_0^q})$ \\
                                       &$2.0^{+1.7}_{-0.8}(m_c)^{+0.2}_{-0.2}(\bar f_S)^{+0.5}_{-0.0}(a_{1,2}^{K^*})^{+0.7}_{-0.3}(B_{1,3}^{S}) (\rm{f_0^s})$ \\
$\rm{B_c \to K^{*+} f_0}$         &$1.5^{+0.4}_{-0.2}(m_c)^{+0.2}_{-0.1}(\bar f_S)^{+0.2}_{-0.1}(a_{1,2}^{K^*})^{+0.5}_{-0.4}(B_{1,3}^{S}) (\rm{f_0^q})$\\
                                       &$1.9^{+1.2}_{-0.8}(m_c)^{+0.2}_{-0.2}(\bar f_S)^{+0.2}_{-0.1}(a_{1,2}^{K^*})^{+0.4}_{-0.4}(B_{1,3}^{S}) (\rm{f_0^s})$\\
\hline \hline
\end{tabular}
\end{center}
\end{table}

As mentioned in Sec.~\ref{sec:1},\;  since the experimental constraints indicate that the mixing angle $\theta_0$ between $\sigma$ and $f_0$
lies in the range of $[25^\circ, 40^\circ]$ or $[140^\circ, 165^\circ]$~\cite{Mixangle:theta0}, then the
pQCD predictions of the BRs for $B_c \to \pi^+ (\sigma, f_0)$ decays with mixing patterns can
be read,
\beq
Br(B_c \to \pi^+ \sigma) &\approx& \left\{ \begin{array}{ll}
(1.9  \sim   2.6)  \times  10^{-7}& {\rm for \ \ 25^\circ< \theta_0< 40^\circ  } \\
 (1.9  \sim   3.0)  \times  10^{-7}& {\rm for \ \ 140^\circ< \theta_0< 165^\circ } \\ \end{array} \right.   \label{eq:pipf60}  \;,\\
Br(B_c \to \pi^+ f_0) &\approx&  \left\{ \begin{array}{ll}
(0.3  \sim   0.8)  \times  10^{-7}& {\rm for \ \ 25^\circ< \theta_0< 40^\circ  } \\
 (0.1  \sim   0.8)  \times  10^{-7}& {\rm for \ \ 140^\circ< \theta_0< 165^\circ } \\ \end{array} \right.  \label{eq:pipf980}\;,
\eeq
where only the central values are quoted, so are the similar cases in the following text unless otherwise stated. Likewise, the
pQCD predictions of the BRs for $B_c \to \rho^+ (\sigma, f_0)$ decays are as follows,
\beq
Br(B_c \to \rho^+ \sigma) &\approx& \left\{ \begin{array}{ll}
(0.9  \sim   1.3)  \times  10^{-7}& {\rm for \ \  25^\circ< \theta_0< 40^\circ  } \\
 (0.9  \sim   1.5)  \times  10^{-7}& {\rm for \ \  140^\circ< \theta_0< 165^\circ } \\ \end{array} \right.   \label{eq:rhopf60}  \;,\\
Br(B_c \to \rho^+ f_0) &\approx&  \left\{ \begin{array}{ll}
(0.1  \sim   0.3)  \times  10^{-7}& {\rm for \ \  25^\circ< \theta_0< 40^\circ  } \\
 (0.05  \sim   0.3)  \times  10^{-7}& {\rm for \ \  140^\circ< \theta_0< 165^\circ } \\ \end{array} \right.  \label{eq:rhopf980}\;.
\eeq
\begin{table}[b]
\caption{ Same as Table~\ref{tab:bcsp-1s0} but for the $\Delta S = 0$ processes of
charmless hadronic $B_c \to (a_0(1450), {K_0^*}(1430), f_0(1370), f_0(1500))(\rho, K^*, \omega, \phi)$ decays
 in S1 and S2, respectively.}
\label{tab:bcsv-2s0}
\begin{center}\vspace{-0.4cm}
\begin{tabular}[t]{l|l}
 \hline \hline
 $\Delta S =0$     &                                            \\
 Decay modes       &    BRs $(10^{-6})$         \\
 \hline
$\rm{B_c \to a_0(1450)^+ \rho^0}$                 &$47.0^{+16.5}_{-12.5}(m_c)^{+11.2}_{-9.4}(\bar f_S)^{+7.3}_{-8.3}(a_2^\rho)^{+13.5}_{-7.5}(B_{1,3}^{S})$ (S1)  \\
                                                  &$15.3^{+11.9}_{-6.3}(m_c)^{+3.5}_{-3.1}(\bar f_S)^{+0.8}_{-0.2}(a_2^\rho)^{+7.8}_{-3.7}(B_{1,3}^{S})$ (S2) \\
$\rm{B_c \to a_0(1450)^0 \rho^+}$            &$27.4^{+8.9}_{-6.4}(m_c)^{+6.2}_{-5.5}(\bar f_S)^{+2.4}_{-3.5}(a_2^\rho)^{+6.3}_{-6.1}(B_{1,3}^{S})$ (S1)  \\
                                             &$15.5^{+9.3}_{-6.4}(m_c)^{+3.5}_{-3.2}(\bar f_S)^{+0.2}_{-0.0}(a_2^\rho)^{+5.8}_{-4.8}(B_{1,3}^{S})$ (S2)
\\
$\rm{B_c \to a_0(1450)^+ \omega}$            &$6.5^{+2.0}_{-1.1}(m_c)^{+1.4}_{-1.4}(\bar f_S)^{+0.6}_{-0.0}(a_2^\omega)^{+2.2}_{-1.8}(B_{1,3}^{S})$ (S1)  \\
                                             &$1.1^{+0.3}_{-0.1}(m_c)^{+0.3}_{-0.2}(\bar f_S)^{+0.2}_{-0.0}(a_2^\omega)^{+3.1}_{-0.7}(B_{1,3}^{S})$ (S2)\\
   \hline
$\rm{B_c \to \overline{K}_0^*(1430)^0 K^{*+}}$             &$35.7^{+18.8}_{-12.3}(m_c)^{+8.3}_{-6.5}(\bar f_S)^{+2.9}_{-3.5}(a_{1,2}^{K^*})^{+4.3}_{-3.1}(B_{1,3}^{S})$ (S1)\\
                                                           &$5.4^{+5.5}_{-2.8}(m_c)^{+1.4}_{-1.0}(\bar f_S)^{+0.7}_{-0.4}(a_{1,2}^{K^*})^{+5.6}_{-1.9}(B_{1,3}^{S})$ (S2)\\
$\rm{B_c \to \overline{K}^{*0} {K_0^*}(1430)^+}$                &$5.0^{+0.9}_{-1.6}(m_c)^{+1.1}_{-0.7}(\bar f_S)^{+1.0}_{-0.8}(a_{1,2}^{K^*})^{+1.6}_{-0.8}(B_{1,3}^{S})$  (S1) \\
                                                                &$8.2^{+2.9}_{-2.0}(m_c)^{+2.0}_{-1.9}(\bar f_S)^{+1.8}_{-1.7}(a_{1,2}^{K^*})^{+6.1}_{-4.0}(B_{1,3}^{S})$  (S2)\\
  \hline
$\rm{B_c \to f_0(1370) \rho^+}$              &$6.1^{+3.9}_{-2.1}(m_c)^{+1.6}_{-1.3}(\bar f_S)^{+2.8}_{-1.8}(a_2^\rho)^{+2.2}_{-1.5}(B_{1,3}^{S}) (\rm{f_0^q, S1})$ \\
                                             &$1.7^{+0.3}_{-0.2}(m_c)^{+0.3}_{-0.4}(\bar f_S)^{+0.5}_{-0.4}(a_2^\rho)^{+1.1}_{-1.3}(B_{1,3}^{S}) (\rm{f_0^q, S2})$    \\
$\rm{B_c \to f_0(1500) \rho^+}$              &$6.1^{+3.7}_{-2.4}(m_c)^{+1.5}_{-1.3}(\bar f_S)^{+2.4}_{-1.8}(a_2^\rho)^{+2.2}_{-1.7}(B_{1,3}^{S}) (\rm{f_0^q, S1})$  \\
                                             &$1.7^{+0.5}_{-0.1}(m_c)^{+0.4}_{-0.3}(\bar f_S)^{+0.6}_{-0.4}(a_2^\rho)^{+0.9}_{-1.4}(B_{1,3}^{S}) (\rm{f_0^q, S2})$
 \\
\hline \hline
\end{tabular}
\end{center}
\end{table}

According to Ref.~\cite{Cheng06:glueball}, $f_0(1370)$ and $f_0(1500)$ mixing has the following form,
\beq
f_0(1370) &=& 0.78 f_0^q + 0.51 f_0^s\;, \quad   f_0(1500) = -0.54 f_0^q +0.84 f_0^s \;,
\eeq
where we neglect the possible small or tiny scalar glueball components in the present paper and leave them
for future study. 
Then the
pQCD predictions of the BRs for $B_c \to \pi^+ (f_0(1370), f_0(1500))$ decays can
be read,
\beq
Br(B_c \to \pi^+ f_0(1370)) &\approx& \left\{ \begin{array}{ll}
2.2  \times  10^{-6} ({\rm S1})\;,&  \\
6.0  \times  10^{-7} ({\rm S2})\;; &   \\ \end{array} \right.     \\
Br(B_c \to \pi^+ f_0(1500)) &\approx&  \left\{ \begin{array}{ll}
1.1  \times  10^{-6} ({\rm S1})\;,&  \\
2.7  \times  10^{-7} ({\rm S2})\;; &   \\ \end{array} \right.
\eeq
\begin{table}[b]
\caption{ Same as Table~\ref{tab:bcsp-1s0} but for the $\Delta S = 1$ processes of
charmless hadronic $B_c \to (a_0(1450), {K_0^*}(1430), f_0(1370), f_0(1500))(\rho, K^*, \omega, \phi)$ decays
 in S1 and S2, respectively.}
\label{tab:bcsv-2s1}
\begin{center}\vspace{-0.6cm}
\begin{tabular}[t]{l|l}
 \hline \hline
$\Delta S =1$ &                      \\
Decay modes   &    BRs $(10^{-7})$    \\
 \hline
$\rm{B_c \to a_0(1450)^+ K^{*0}}$             &$2.7^{+0.4}_{-0.6}(m_c)^{+0.5}_{-0.5}(\bar f_S)^{+0.1}_{-0.3}(a_{1,2}^{K^*})^{+0.9}_{-0.6}(B_{1,3}^{S})$ (S1) \\
                                              &$7.0^{+1.9}_{-1.7}(m_c)^{+1.7}_{-1.6}(\bar f_S)^{+1.2}_{-0.8}(a_{1,2}^{K^*})^{+1.8}_{-3.5}(B_{1,3}^{S})$ (S2) \\
$\rm{B_c \to a_0(1450)^0 K^{*+}}$            &$1.4^{+0.2}_{-0.3}(m_c)^{+0.3}_{-0.3}(\bar f_S)^{+0.1}_{-0.1}(a_{1,2}^{K^*})^{+0.5}_{-0.3}(B_{1,3}^{S})$ (S1) \\
                                             &$3.5^{+1.0}_{-0.9}(m_c)^{+0.9}_{-0.8}(\bar f_S)^{+0.6}_{-0.4}(a_{1,2}^{K^*})^{+0.9}_{-1.8}(B_{1,3}^{S})$ (S2)
\\
  \hline
$\rm{B_c \to {K_0^*}(1430)^+ \omega}$        &$7.1^{+3.3}_{-2.7}(m_c)^{+1.5}_{-1.3}(\bar f_S)^{+0.7}_{-0.8}(a_2^\omega)^{+0.8}_{-0.8}(B_{1,3}^{S})$ (S1) \\
                                             &$1.3^{+1.2}_{-0.6}(m_c)^{+0.3}_{-0.2}(\bar f_S)^{+0.1}_{-0.1}(a_2^\omega)^{+1.0}_{-0.3}(B_{1,3}^{S})$ (S2) \\
$\rm{B_c \to {K_0^*}(1430)^+ \phi}$          &$2.8^{+0.4}_{-0.6}(m_c)^{+0.8}_{-0.5}(\bar f_S)^{+0.3}_{-0.6}(a_2^\phi)^{+0.8}_{-0.5}(B_{1,3}^{S})$ (S1) \\
                                             &$4.3^{+1.7}_{-1.3}(m_c)^{+1.2}_{-1.0}(\bar f_S)^{+0.7}_{-0.4}(a_2^\phi)^{+3.8}_{-2.1}(B_{1,3}^{S})$ (S2)
 \\
   \hline
$\rm{B_c \to {K_0^*}(1430)^0 \rho^+}$            &$16.5^{+8.9}_{-5.9}(m_c)^{+3.2}_{-3.1}(\bar f_S)^{+0.9}_{-2.0}(a_2^\rho)^{+1.6}_{-1.8}(B_{1,3}^{S})$ (S1) \\
                                                 &$2.9^{+3.1}_{-1.6}(m_c)^{+0.8}_{-0.5}(\bar f_S)^{+0.2}_{-0.0}(a_2^\rho)^{+2.6}_{-0.8}(B_{1,3}^{S})$ (S2) \\
$\rm{B_c \to {K_0^*}(1430)^+ \rho^0}$        &$8.2^{+4.5}_{-3.0}(m_c)^{+1.6}_{-1.6}(\bar f_S)^{+0.5}_{-1.0}(a_2^\rho)^{+0.8}_{-0.9}(B_{1,3}^{S})$ (S1) \\
                                             &$1.5^{+1.5}_{-0.8}(m_c)^{+0.4}_{-0.3}(\bar f_S)^{+0.1}_{-0.0}(a_2^\rho)^{+1.3}_{-0.4}(B_{1,3}^{S})$ (S2) \\
  \hline
$\rm{B_c \to f_0(1370) K^{*+}}$              &$1.0^{+0.3}_{-0.2}(m_c)^{+0.3}_{-0.2}(\bar f_S)^{+0.2}_{-0.0}(a_{1,2}^{K^*})^{+0.6}_{-0.1}(B_{1,3}^{S}) (\rm{f_0^q, S1})$ \\
                                             &$2.4^{+0.8}_{-0.5}(m_c)^{+0.6}_{-0.4}(\bar f_S)^{+0.6}_{-0.2}(a_{1,2}^{K^*})^{+1.4}_{-1.4}(B_{1,3}^{S}) (\rm{f_0^q, S2})$ \\
                                               &$14.3^{+5.8}_{-4.0}(m_c)^{+3.5}_{-3.2}(\bar f_S)^{+2.8}_{-3.5}(a_{1,2}^{K^*})^{+3.1}_{-3.3}(B_{1,3}^{S}) (\rm{f_0^s, S1})$ \\
                                               &$3.2^{+2.5}_{-1.3}(m_c)^{+0.7}_{-0.6}(\bar f_S)^{+0.5}_{-0.4}(a_{1,2}^{K^*})^{+2.3}_{-1.0}(B_{1,3}^{S}) (\rm{f_0^s, S2})$ \\
$\rm{B_c \to f_0(1500) K^{*+}}$            &$1.1^{+0.2}_{-0.3}(m_c)^{+0.3}_{-0.2}(\bar f_S)^{+0.0}_{-0.1}(a_{1,2}^{K^*})^{+0.4}_{-0.3}(B_{1,3}^{S}) (\rm{f_0^q, S1})$ \\
                                           &$2.5^{+0.7}_{-0.7}(m_c)^{+0.5}_{-0.5}(\bar f_S)^{+0.4}_{-0.3}(a_{1,2}^{K^*})^{+1.1}_{-1.6}(B_{1,3}^{S}) (\rm{f_0^q, S2})$ \\
                                               &$14.9^{+5.6}_{-5.0}(m_c)^{+3.7}_{-3.3}(\bar f_S)^{+2.3}_{-3.9}(a_{1,2}^{K^*})^{+2.4}_{-3.7}(B_{1,3}^{S}) (\rm{f_0^s, S1})$\\
                                               &$3.5^{+2.6}_{-1.5}(m_c)^{+0.8}_{-0.7}(\bar f_S)^{+0.4}_{-0.5}(a_{1,2}^{K^*})^{+2.2}_{-1.1}(B_{1,3}^{S}) (\rm{f_0^s, S2})$ \\
\hline \hline
\end{tabular}
\end{center}
\end{table}
Likewise, the
pQCD predictions of the BRs for $B_c \to \rho^+ (f_0(1370), f_0(1500))$ decays are 
\beq
Br(B_c \to \rho^+ f_0(1370)) &\approx& \left\{ \begin{array}{ll}
3.7  \times  10^{-6} ({\rm S1})\;,&  \\
1.0  \times  10^{-6} ({\rm S2})\;; &   \\ \end{array} \right.     \\
Br(B_c \to \rho^+ f_0(1500)) &\approx&  \left\{ \begin{array}{ll}
1.8  \times  10^{-6} ({\rm S1})\;,&  \\
5.0  \times  10^{-7} ({\rm S2})\;. &  \\ \end{array} \right.
\eeq

For the $\Delta S =1$ processes $B_c \to K^{(*)+} (f, f')$ decays,
the BRs in the pQCD approach based on the pure $q \bar q$ state $f_0^q$ or
pure $s\bar s$ one $f_0^s$ of the scalars $f$ and $f'$ are given in the Tables~\ref{tab:bcsp-1s1}, \ref{tab:bcsp-2s1},
\ref{tab:bcsv-1s1}, and \ref{tab:bcsv-2s1}. One can observe straightforwardly from the tables that
all the BRs for $B_c \to  K^{(*)+} (f, f')$ channels are in the order of $10^{-8} \sim 10^{-7}$
except for $B_c \to K^{*+} f_0(1500)$ in S1 though which is CKM suppressed. But, if the branching ratio
of $10^{-6}$ for $B_c \to K^{*+} f_0(1500)$ decay can be detected by
the experiments, it is doubtless that
the scalar meson $f_0(1500)$ is dominated by the $s\bar s$ component.
When we consider the mixing form for the scalars $f$ and $f'$, the {\it CP}-averaged BRs for $B_c \to K^{(*)+} (f, f')$ decays
within the pQCD approach have been calculated and
shown in the Eqs.~(\ref{eq:kpf60}-\ref{eq:kspf150}):
\beq
Br(B_c \to K^{+}  \sigma)  &\approx&  \left\{ \begin{array}{ll}
(2.0  \sim   2.0)  \times  10^{-7}& {\rm for \ \  25^\circ< \theta_0< 40^\circ  } \\
 (0.5  \sim   1.1)  \times  10^{-7}& {\rm for \ \  140^\circ< \theta_0< 165^\circ } \\ \end{array} \right.  \label{eq:kpf60} \;,\\
Br(B_c \to K^{+}  f_0)  &\approx&  \left\{ \begin{array}{ll}
(0.2  \sim   0.5)  \times  10^{-7}& {\rm for \ \  25^\circ< \theta_0< 40^\circ   } \\
 (0.6  \sim   1.4)  \times  10^{-7}& {\rm for \ \  140^\circ< \theta_0< 165^\circ } \\ \end{array} \right.  \label{eq:kpf980} \;;
\eeq
\beq
Br(B_c \to K^{*+}  \sigma)  &\approx&  \left\{ \begin{array}{ll}
(3.0  \sim   3.5)  \times  10^{-7}& {\rm for \ \  25^\circ< \theta_0< 40^\circ  } \\
 (0.06  \sim   0.8)  \times  10^{-7}& {\rm for \ \  140^\circ< \theta_0< 165^\circ } \\ \end{array} \right.  \label{eq:kspf60} \;,\\
Br(B_c \to K^{*+}  f_0)  &\approx&  \left\{ \begin{array}{ll}
(0.1  \sim   0.6)  \times  10^{-7}& {\rm for \ \  25^\circ< \theta_0< 40^\circ  } \\
 (2.7  \sim   3.4)  \times  10^{-7}& {\rm for \ \  140^\circ< \theta_0< 165^\circ } \\ \end{array} \right.  \label{eq:kspf980} \;;
\eeq
\beq
Br(B_c \to K^{+}  f_0(1370))  &\approx&  \left\{ \begin{array}{ll}
1.4  \times  10^{-7} ({\rm S1})\;,&  \\
1.8  \times  10^{-7} ({\rm S2})\;; &   \\ \end{array} \right.    \\
Br(B_c \to K^{+}  f_0(1500))  &\approx& \left\{ \begin{array}{ll}
7.1  \times  10^{-7} ({\rm S1})\;,&  \\
1.3  \times  10^{-7} ({\rm S2})\;; &  \\ \end{array} \right.
\eeq
\beq
Br(B_c \to K^{*+}  f_0(1370))  &\approx&  \left\{ \begin{array}{ll}
1.8  \times  10^{-7} ({\rm S1})\;,&  \\
6.3  \times  10^{-8} ({\rm S2})\;; &   \\ \end{array} \right.    \\
Br(B_c \to K^{*+}  f_0(1500))  &\approx& \left\{ \begin{array}{ll}
1.4  \times  10^{-6} ({\rm S1})\;,&  \\
5.2  \times  10^{-7} ({\rm S2})\;. &   \\ \end{array} \right. \label{eq:kspf150}
\eeq

Hence, based on the numerical results shown in the Tables~\ref{tab:bcsp-1s0}, \ref{tab:bcsp-1s1},
\ref{tab:bcsv-1s0}, and \ref{tab:bcsv-1s1}, and Eqs.~(\ref{eq:pipf60}-\ref{eq:rhopf980}) and~(\ref{eq:kpf60}-\ref{eq:kspf980}),
it is evident that
 the theoretical implications on the components
of $\sigma$ and $f_0$ in the light scalar nonet can not be provided by the small pQCD predictions on
the short-distance contributions of $B_c \to (\pi^+, K^+, \rho^+, K^{*+}) (\sigma, f_0)$ decays.
However, once the large BRs above $10^{-6}$ for $\Delta S =0$ processes
$B_c \to (\pi^+, \rho^+) (f_0(1370), f_0(1500))$ in both scenarios and $\Delta S= 1$ $B_c \to K^{*+} f_0(1500)$ decay in
scenario 1 could be measured in the ongoing LHCb or the forthcoming Sper-B experiments,
they may help determine the
components,
the ratios of quarkonia, and
the preferred scenario by the experiments for these two considered scalar $f_0(1370)$ and $f_0(1500)$ mesons, respectively.


\vspace{0.8cm}

Frankly speaking, for many considered pure annihilation $B_c$ decays with BRs of or below $10^{-7}$, it is still hard to observe them
even in LHC due to their tiny decay rates.
Their observation at LHC, however, would mean a large non-perturbative
contribution or a signal for exotic new physics beyond the SM.
It is worth of stressing that the theoretical predictions in
the pQCD approach still have large theoretical errors induced
by the still large uncertainties of many input parameters.
Any progress in reducing the error of input parameters, such as
the Gegenbauer moments $a_i$ of the pseudoscalar or vector mesons distribution amplitudes,
$B_i$ of the scalar mesons distribution amplitudes and the charm quark mass $m_c$, will help us
to improve the precision of the pQCD predictions.  We do not consider the possible long-distance
contributions, such as the rescattering effects, although they should be present, and they
may be large and affect the theoretical predictions. It is beyond
the scope of this work and expected to be studied in the future work.


\section{Summary}\label{sec:sum}

In summary, we studied the two-body charmless hadronic $B_c \to SP, SV$
decays by employing the pQCD factorization approach based on the
$k_T$ factorization theorem. These considered decay channels can
occur only via the annihilation diagrams and they will provide an important
testing ground for the magnitude of the annihilation contributions and implications
to the mechanism of annihilation decays. Based on the assumption of
two-quark structure of the light scalars, we make the theoretical predictions
on the {\it CP}-averaged branching ratios of considered $B_c \to SP, SV$ channels. In turn,
we could obtain the implications on the component and physical properties of the light scalar mesons
through the experimental measurements on these considered charmless hadronic $B_c$ decays.
Furthermore, these decay modes might also reveal the existence of exotic new physics
scenario or nonperturbative QCD effects.

The pQCD predictions for {\it CP}-averaged branching ratios are
displayed in Tables~\ref{tab:bcsp-1s0}-\ref{tab:bcsv-2s1}.
From our numerical evaluations and phenomenological analysis, we
found the following results:
\begin{itemize}
\item
The pQCD predictions for the branching ratios
vary in the range of $10^{-5}$ to $10^{-8}$.
Many decays with a decay rate
at $10^{-6}$ or larger could be measured at the LHCb experiment.

\item
For $B_c \to SP, SV$ decays, the branching ratios of  $\Delta S= 0$ processes
are basically larger than those of $\Delta S =1$ ones.
Such differences are mainly induced  by
the CKM factors involved: $V_{ud}\sim 1 $ for the former decays
while $V_{us}\sim 0.22$ for the latter ones.

\item
Analogous to $B \to K^* \eta^{(\prime)}$ decays, we find
$Br(B_c \to \kappa^+ \eta) \sim 5 \times Br(B_c \to \kappa^+ \eta')$.
This 
difference can be understood
by the destructive and constructive interference
between the $\eta_q$ and $\eta_s$ contribution to
the $B_c \to \kappa^+ \eta'$ and $B_c \to \kappa^+ \eta$ decay, respectively.

\item
For $B_c \to K_0^*(1430) \eta^{(')}$ channels, the branching ratios for these two decays
are similar to each other in both scenarios, which is mainly because the factorizable
contributions of $\eta_s$ term play the dominant role and expected to be tested by the
forthcoming Super-B experiments.

\item
If $a_0$ and $\kappa$ are the $q\bar q$ bound states, the pQCD predicted
BRs for $B_c \to a_0 (\pi, \rho)$ and $B_c \to \kappa K^{(*)}$ decays will be in the range of
$10^{-6} \sim 10^{-5}$, which are within the reach of the LHCb experiments and expected to be measured.

\item
For the $a_0(1450)$ and $K_0^*(1430)$ channels, the BRs for $B_c \to a_0(1450) (\pi, \rho)$
and $B_c \to K_0^*(1430) K^{(*)}$ modes in the pQCD approach are found to be of order $(5 \sim 47) \times 10^{-6}$
and $(0.7 \sim 36) \times 10^{-6}$, respectively. A measurement of them at the predicted level will
favor the structure of $q\bar q$ for the $a_0(1450)$ and $K_0^*(1430)$ and identify which scenario
is preferred.

\item
Because only tree operators are involved, the {\it CP}-violating asymmetries for
these considered $B_c$ decays are absent naturally.

\item
The pQCD predictions still have large theoretical uncertainties,
induced by the uncertainties of input parameters.

\item
We here calculated the branching ratios of the pure annihilation $B_c \to SP, SV$ decays by employing the pQCD approach.
We do not consider the possible long-distance contributions, such as the
re-scattering effects, although they may be large and affect the theoretical
predictions. It is beyond the  scope of this work.

\end{itemize}

\begin{acknowledgments}

X.~Liu would like to thank Professors Cai-dian~L\"u and Hsiang-nan~Li for valuable discussions
and Dr. You-chang~Yang for reading the manuscript.
This work is supported by the National Natural Science Foundation of China
under Grant No.10975074, and No.10735080, by the Project on Graduate
Students' Education and Innovation of Jiangsu
Province, under Grant No. ${\rm CX09B_{-}297Z}$,  and by the Project on
Excellent Ph.D Thesis of Nanjing Normal University, under Grant
No. 181200000251.

\end{acknowledgments}




\begin{thebibliography}{99}
\bibitem{Amsler08:pdg}
 S.~Spanier, N.A.~T\"ornqvist, and C.~Amsler, (Particle Data Group), ``Note on scalar mesons";
 C.~Amsler {\it et al.}, (Particle Data Group), \plb {\bf 667}, 1 (2008).

\bibitem{Godfrey99:scalar}
 S.~Godfrey and J.~Napolitano, \rmp {\bf 71}, 1411 (1999).

\bibitem{Close02:scalar}
 F.E.~Close and N.A.~T\"ornqvist, \jpg {\bf 28}, R249 (2002).

\bibitem{belle02:f98}
A.~Garmash {\it et al.}, (Belle Collaboration), \prd {\bf 65}, 092005 (2002).

\bibitem{belle05:f98}
A.~Garmash {\it et al.}, (Belle Collaboration), \prd {\bf 71}, 092003 (2005);
K.~Abe {\it et al.}, (Belle Collaboration),
arXiv:0509001[hep-ex]; J.~Dragic, talk presented at the HEP2005
Europhysics Conference in Lisboa, Portugal, July 21-27, 2005.

\bibitem{babar04:f98}
B.~Aubert {\it et al.}, (BaBar Collaboration), \prd {\bf 70}, 092001 (2004).

\bibitem{Barberio08:hfag}
Heavy Flavor Averaging Group, E.~Barberio {\it et al.,}
arXiv:0808.1297[hep-ex]; and online update at
http://www.slac.stanford.edu/xorg/hfag.

\bibitem{Chen:b2scalar}
C.H.~Chen, \prd {\bf 67}, 014012 (2003);
ibid. {\bf 67}, 094011 (2003);
C.H.~Chen and C.Q.~Geng, \prd {\bf 75}, 054010 (2007).

\bibitem{Cheng:b2scalar}
H.Y.~Cheng and K.C.~Yang, \prd {\bf 71}, 054020 (2005);
P.~Minkowski and W.~Ochs, \epjc {\bf 39}, 71 (2005);
A. K. Giri, B. Mawlong, and R. Mohanta, \prd {\bf 74}, 114001 (2006).


\bibitem{Cheng06:B2SP}
H.Y.~Cheng, C.K.~Chua, and K.C.~Yang, \prd {\bf 73}, 014017 (2006), and reference
therein;
H.Y.~Cheng, C.K.~Chua, and K.C.~Yang, \prd {\bf 77}, 014034 (2008).

\bibitem{Wang:b2scalar}
W.~Wang, Y.L.~Shen, Y.~Li, and C.D.~L\"u, \prd {\bf 74}, 114010 (2006);
Y.L.~Shen, W.~Wang, J.~Zhu, and C.D.~L\"u, \epjc {\bf 50}, 877 (2007);
C.S.~Kim, Y.~Li, and W.~Wang, \prd {\bf 81}, 074014 (2010);
P.~Colangelo, F.DeFazio, and W.~Wang, \prd {\bf 81}, 074001 (2010).

\bibitem{Xiao:b2scalar}
X.~Liu and Z.J.~Xiao, \ctp {\bf 53}, 540 (2010);
X.~Liu, Z.Q.~Zhang, and Z.J.~Xiao, \cpc {\bf 34}, 157 (2010);
Z.Q.~Zhang and Z.J.~Xiao, \cpc {\bf 34}, 528 (2010);
arXiv:0812.2314 [hep-ph].

\bibitem{nb04:bcre}
N.~Brambilla {\it et al.}, (Quarkonium Working Group),
CERN-2005-005, arXiv:0412158[hep-ph].

\bibitem{Lu03:dsk}
C.D.~L\"u and K.~Ukai, \epjc {\bf 28}, 305 (2003).

\bibitem{Keum01:kpi}
Y.Y.~Keum, H.N.~Li and A.I.~Sanda, \plb {\bf 504}, 6 (2001);
\prd {\bf 63}, 054008 (2001).


\bibitem{Lu01:pipi}
C.D.~L\"u, K.~Ukai and M.Z.~Yang,  \prd {\bf 63}, 074009 (2001).

\bibitem{Hong06:direct}
B.H.~Hong and C.D.~L\"u,  Sci.\ China  G {\bf 49}, 357 (2006).


\bibitem{Li05:kphi}
H.N.~Li, and S.~Mishima, \prd {\bf 71}, 054025 (2005);
H.N.~Li, \plb {\bf 622}, 63 (2005).

\bibitem{Gritsan07:kphi}
A.V.~Gritsan,  Econf. C {\bf 070512}, 001 (2007). 
A.L.~Kagan, \plb {\bf 601}, 151 (2004).

\bibitem{Stewart08:anni-scet}
C.M.~Arnesen, Z.~Ligeti, I.Z.~Rothstein, and I.W.~Stewart, \prd{\bf 77}, 054006 (2008).

\bibitem{scet}
C.W.~Bauer, S.~Fleming, and M.E.~Luke, \prd {\bf 63}, 014006 (2000);
C.W.~Bauer, S.~Fleming, D.~Pirjol, and I.W.~Stewart, \prd {\bf 63}, 114020 (2001);
C.W.~Bauer and I.W.~Stewart, \plb {\bf 516}, 134 (2001);
C.W.~Bauer, D.~Pirjol, and I.W.~Stewart,\prd {\bf 65}, 054022 (2002);
C.W.~Bauer, S.~Fleming, D.~Pirjol, I.Z.~Rothstein, and I.W.~Stewart, \prd {\bf 66}, 014017 (2002).

\bibitem{Chay08:complexanni}
J.~Chay, H.N.~Li, and S.~Mishima, \prd{\bf 78}, 034037, (2008).


\bibitem{Buras96:weak}
G.~Buchalla, A.J.~Buras and M.E.~Lautenbacher, \rmp {\bf 68}, 1125 (1996).

\bibitem{Li03:ppnp}
H.N.~Li, Prog. Part. $\&$ Nucl. Phys. {\bf 51}, 85 (2003), and reference
therein.

\bibitem{qcdf}
M.~Beneke, G.~Buchalla, M.~Neubert and C.T.~Sachrajda, \prl  {\bf
83}, 1914 (1999); \npb {\bf 591}, 313 (2000).

\bibitem{Li09:review}
H.N.~Li, arXiv:0907.4940[hep-ph].

\bibitem{xiao06}
X.~Liu, H.S.~Wang, Z.J.~Xiao, L.B.~Guo, and C.D.~L\"u, \prd {\bf 73}, 074002 (2006);
H.S.~Wang, X.~Liu, Z.J.~Xiao, L.B.~Guo, and C.D.~L\"u, \npb {\bf 738}, 243 (2006);
Z.J.~Xiao, X.F.~Chen,  and D.Q.~Guo, \epjc {\bf 50}, 363 (2007);
Z.J.~Xiao, D.Q~Guo,  and X.F.~Chen, \prd {\bf 75}, 014018 (2007);
Z.J.~Xiao, X.~Liu, and H.S.~Wang, \prd {\bf 75}, 034017 (2007);
Z.J.~Xiao, X.F.~Chen,  and D.Q.~Guo, arXiv:0701146[hep-ph].

\bibitem{Ali07:bsnd}
A.~Ali, G.~Kramer, Y.~Li, C.D.~L\"u, Y.L.~Shen, W.~Wang, and Y.M.~Wang, \prd {\bf 76}, 074018 (2007);

\bibitem{xiao08:keta}
Z.J.~Xiao, Z.Q.~Zhang, X.~Liu, and L.B.~Guo, \prd {\bf 78}, 114001 (2008).


\bibitem{Xiao1:bcdecay}
X.~Liu, Z.J.~Xiao, and C.D.~L\"u, \prd {\bf 81}, 014022 (2010);
X.~Liu and Z.J.~Xiao, \prd {\bf 81}, 074017 (2010);
X.~Liu and Z.J.~Xiao, arXiv:1003.3929[hep-ph].

\bibitem{Mixangle:theta0}
 M.~Alford and R.L.~Jaffe, \npb {\bf 578}, 367 (2000);
 H.Y.~Cheng, \prd {\bf 67}, 034024 (2003);
 A.V.~Anisovich, V.V.~Anisovich, and V.A.~Nikonov, Eur. Phys. J. A {\bf 12}, 103 (2001);
 Phys. At. Nucl. {\bf 65}, 497 (2002);
 A.~Gokalp, Y.~Sarac, and O.~Yilmaz, \plb {\bf 609}, 291 (2005).


\bibitem{Cheng06:glueball}
H.Y.~Cheng, C.K.~Chua, and K.F.~Liu, \prd {\bf 74}, 094005 (2006).

\bibitem{Li02:resum}
H.N.~Li, \prd {\bf 66}, 094010 (2002).

\bibitem{Li98:soft}
H.N.~Li and B.~Tseng,\prd {\bf 57}, 443 (1998).

\bibitem{luy03:form}
C.D.~L\"u and M.Z.~Yang, \epjc {28}, 515 (2003).

\bibitem{CDL}
J.F.~Cheng, D.S.~Du and C.D.~L\"{u}, \epjc {\bf 45}, 711 (2006).

\bibitem{Li09:B2Sfm}
R.H.~Li, C.D.~L\"u, W.~Wang, and X.X.~Wang, \prd {\bf 79}, 014013 (2009)

\bibitem{ekou09:ncbc}
S.~Descotes-Genon, J.~He, E.~Kou, and P.~Robbe, \prd {\bf 80}, 114031 (2009).

\bibitem{Mathur06:lattice}
N.~Mathur, A.~Alexandru, Y.~Chen, S.J.~Dong, T.~Draper, I.~Horv\'ath, F.X.~Lee, K.F.~Liu, S.~Tamhankar, and J.B.~Zhang, \prd {\bf 76}, 114505 (2007).

\bibitem{Kim97-Burch06:lattice}
W.~Lee and D.~Weingarten, \prd {\bf 61}, 014015 (1999);
M.~G\"ockeler, R.~Horsley, H.~Perlt, P.~Rakow, G.~Schierholz, A.~Schiller, and P.~Stephenson, \prd {\bf 57}, 5562 (1998);
S.~Kim and S.~Ohta, Nucl. Phys. Proc. Suppl. B {\bf 53}, 199 (1997);
A.~Hart, C.~McNeile, and C.~Michael, Nucl. Phys. Proc. Suppl. B {\bf 119}, 266 (2003);
T.~Burch, C.~Gattringer, L.Y.~Glozman, C.~Hagen, C.B.~Lang, and A.~Sch\"afer, (BGR [Bern-Graz-Regensburg] Collaboration), \prd {\bf 73}, 094505 (2006).

\bibitem{Bardeen02:lattice}
W.~Bardeen, A.~Duncan, E.~Eichten, N.~Isgur, and H.~Thacker, \prd {\bf 65}, 014509 (2001).

\bibitem{Kunihiro04:lattice}
T.~Kunihiro, S.~Muroya, A.~Nakamura, C.~Nonaka, M.~Sekiguchi, and H.~Wada, (SCALAR Collaboration), \prd {\bf 70}, 034504 (2004).

\bibitem{Prelovsek04:lattice}
S.~Prelovsek, C.~Dawson, T.~Izubuchi, K.~Orginos, and A.~Soni, \prd {\bf 70}, 094503 (2004).




\end{thebibliography}
\end{document}